\documentclass[journal, twocolumn, 10pt]{IEEEtran}

\usepackage[T1]{fontenc}
\usepackage{lipsum}
\usepackage{blindtext}
\usepackage{multicol}
\usepackage{graphicx}
\usepackage{cite}
\usepackage{amssymb,amsmath}
\usepackage{algorithm}
\usepackage{algpseudocode}
\usepackage{flushend}
\usepackage{multirow}
    \usepackage{lgrind}
\usepackage{hhline}
\usepackage{tcolorbox}
\usepackage{graphicx}
\usepackage{cite}
\usepackage{amssymb,amsmath}
\usepackage{algorithm}
\usepackage{color,soul,colortbl}
\usepackage{cleveref}
\usepackage{subfigure}
\usepackage{booktabs}
\usepackage{xcolor}
\usepackage{graphicx}
\usepackage[export]{adjustbox}
\usepackage{caption}
\usepackage{multirow}
\usepackage{hhline}
\usepackage{psfrag}

\usepackage{widetext}
\usepackage{flushend}
\usepackage{cuted}
\usepackage{float}
\usepackage{lipsum}

\newcounter{tempEquationCounter}
\newcounter{thisEquationNumber}

\usepackage[compact]{titlesec}         
\titlespacing{\section}{0pt}{0pt}{0pt} 
\AtBeginDocument{
  \setlength\abovedisplayskip{0pt}
  \setlength\belowdisplayskip{0pt}}

\usepackage{amsthm}

\theoremstyle{plain}

\theoremstyle{plain}

\theoremstyle{plain}

\providecommand{\lemmaname}{Lemma}
\providecommand{\propositionname}{Proposition}
\providecommand{\theoremname}{Theorem}
\providecommand{\lemmaname}{Lemma}
\providecommand{\propositionname}{Proposition}
\providecommand{\theoremname}{Theorem}
\makeatother
\providecommand{\lemmaname}{Lemma}
\providecommand{\propositionname}{Proposition}
\providecommand{\theoremname}{Theorem}

\definecolor{G}{gray}{0.9}
\definecolor{LC}{rgb}{0.88,1,1}

\usepackage[nolist]{acronym}
\begin{document}
\title{{\LARGE{Optimizing Air-borne Network-in-a-box Deployment for \\ Efficient Remote Coverage }}}

\author{{ Sidrah Javed,~\IEEEmembership{Member, IEEE}, Yunfei Chen,~\IEEEmembership{Senior Member, IEEE}, \\ Mohamed-Slim Alouini, ~\IEEEmembership{Fellow, IEEE}, and Cheng-Xiang Wang, ~\IEEEmembership{Fellow, IEEE} } 
 \thanks{S. Javed and Y. Chen are with the Department of Engineering, University of Durham, DH1 3LE, England. E-mail: \{sidrah.javed, yunfei.chen\} @durham.ac.uk,
 M.S. Alouini is with CEMSE Division,  King Abdullah University of Science and Technology (KAUST), Thuwal, Makkah Province,23955-6900 Saudi Arabia. E-mail:slim.alouini@kaust.edu.sa, and  C.-X. Wang is with the National Mobile Communications Research Laboratory, School of Information Science and Engineering, Southeast University, Nanjing 211189, China. E-mail: chxwang@seu.edu.cn.}}

 \maketitle
 \begin{acronym}
 \acro{UA}{user association}
  \acro{NIB}{network-in-a-box}
  \acro{UAV}{unmanned aerial vehicle}
  \acro{SIC}{successive interference cancellation}
    \acro{HAPS}{high-altitude platform station}
    \acro{NOMA}{non-orthogonal multiple access}
   \acro{LAPS}{low-altitude platform station}
  \acro{SCA}{successive convex approximation}
  \acro{GU}{ground user}
\acro{6G}{sixth generation}
  \acro{RAT}{radio access technology}
\acro{RZF}{regularized zero-forcing}
\acro{SWaP}{size, weight, and power}
\acro{BER}{bit error rate}
\acro{AWGN}{additive white Gaussian noise}
\acro{CDF}{cumulative distribution function}
\acro{CSI}{channel state information}
\acro{FDR}{full-duplex relaying}
\acro{HDR}{half-duplex relaying}
\acro{IC}{interference channel}
\acro{IGS}{improper Gaussian signaling}
\acro{MHDF}{multi-hop decode-and-forward}
\acro{SIMO}{single-input multiple-output}
\acro{MIMO}{multiple-input multiple-output}
\acro{MISO}{multiple-input single-output}
\acro{MRC}{maximum ratio combining}
\acro{PDF}{probability density function}
\acro{PGS}{proper Gaussian signaling}
\acro{RSI}{residual self-interference}
\acro{RV}{random vector}
\acro{r.v.}{random variable}
\acro{HWD}{hardware distortion}
\acro{cHWD}[HWD]{Hardware distortion}
\acro{AS}{asymmetric signaling}
\acro{GS}{geometric shaping}
\acro{PS}{probabilistic shaping}
\acro{HS}{hybrid shaping}
\acro{SISO}{single-input single-output}
\acro{QAM}{quadrature amplitude modulation}
\acro{PAM}{pulse amplitude modulation}
\acro{PSK}{phase shift keying}
\acro{DoF}{degrees of freedom}
\acro{ML}{maximum likelihood}
\acro{MAP}{maximum a posterior}
\acro{SNR}{signal-to-noise ratio}
\acro{SCP}{successive convex programming}
\acro{RF}{radio frequency}
\acro{CEMSE}{Computer, Electrical, and Mathematical Sciences and Engineering}
\acro{KAUST}{King Abdullah University of Science and Technology}
\acro{DM}{distribution matching}
\end{acronym}
\vspace{-10cm}
\begin{abstract}
Among many envisaged drivers for \ac{6G}, one is from the United Nation’s Sustainability Development Goals 2030 to eliminate digital inequality. Remote coverage in sparsely populated areas, difficult terrains or emergency scenarios requires on-demand access and flexible deployment with minimal capex and opex. In this context, \ac{NIB} is an exciting solution which packs the whole wireless network into a single portable and re-configurable box to support multiple access technologies such as WiFi/2G/3G/4G/5G etc. In this paper, we propose \ac{LAPS} based \ac{NIB}s with stratospheric high-altitude platform station \ac{HAPS} as backhaul. Specifically, backhaul employs \ac{NOMA} with superposition coding at the transmitting HAPS and \ac{SIC} at the receiving \ac{NIB}s, whereas the access link (AL) employs superposition coding along with the \ac{RZF} precoding at the \ac{NIB} in order to elevate the computational overhead from the ground users. The required number of airborne \ac{NIB}s to serve a desired coverage area, their optimal placement, user association, beam optimization, and resource allocation are optimized by maximizing the sum-rate of the AL while maintaining the quality of service. Our findings reveal the significance of thorough system planning and communication parameters optimization for enhanced system performance and best coverage under limited resources.
\end{abstract}
\begin{IEEEkeywords}
 HAPS, non-orthogonal multiple access, network-in-a-box, unmanned aerial vehicles, and 6G. 
\end{IEEEkeywords}

%
%

\section{Introduction} 

The upcoming generations of wireless communications envision a comprehensive network capable of delivering ubiquitous and resilient connections to eliminate the digital divide \cite{matracia2023bridging,xu2023toward,wang2023complete}. The emerging technologies, such as aerial communications,  have the potential to reach the unconnected people in remote areas. These platforms, such as drones, balloons, aircrafts, and airships, offer unique advantages that can overcome limitations of traditional terrestrial and satellite communications \cite{chandrasekharan2016designing,wang2023road}. 
For instance, aerial platforms offer flexible deployment rendering rapid on-demand coverage with mobility in remote or disaster-affected areas \cite{zhang2023joint,chang2023novel}. Moreover, they are promising candidates for extensive coverage to reach communities in geographically challenging locations.
In addition, they may be less susceptible to disruption caused by natural disasters compared with ground-based infrastructure. These features make aerial communications a valuable tool for connecting unserved or underserved populations. They can function synergistically with the existing ground and space infrastructure \cite{benzaghta2022uav}, thereby bridging the digital divide and connecting the unconnected.

In aerial communications, multiple platforms can be combined in multi-layer, where \ac{LAPS} can offer access link (AL) to the \ac{GU}s whereas \ac{HAPS} provides the backhaul links to all the serving LAPS.
 HAPS can be implemented in the form of airship, aircraft or tethered balloons, while,
LAPS can be realized as \ac{UAV}s.  {Although HAPS are capable of directly connecting to the GUs using 4G LTE or 5G NR \cite{belmekki2024cellular}, but, the spectrum compatible with the existing GU equipment offers low throughput. On the other hand, employing LAPS as an intermediate layer between HAPS and GUs can increase the overall system throughput by allowing the use of higher frequency bands for backhaul and higher area throughput of LAPS. }

 {The communication networks for aerial platforms must be adaptive and seamlessly integrable. In this context, \ac{NIB} (portable network or pop-out network) emerges as a complete hardware and software solution in a compact reconfigurable package to provide on-demand access while supporting multiple \ac{RAT}s. With efficient \ac{SWaP} characteristics, \ac{NIB}s can become the building blocks for generating flexible and adaptable networks \cite{pozza2018network}.  \ac{NIB}s render widespread applications in commercial, government, and  private sectors for remote coverage, disaster relief or enhanced cyber security \cite{hsu2021guest,anjum2023securitisation}. The concept of a small portable network with few physical devices was conceived in the preliminary works in \cite{sanchez1998rdrn} and \cite{evans1999rapidly}. \ac{NIB}s can either operate as a stand-alone network or co-exist with an exiting network. In this regard, Huang et al. \cite{huang2012design} proposed a physical device that connects to the pre-existent base stations in order to restore the original mobile network. The stand-alone \ac{NIB}-based network was investigated to support smart health IoT services and broadband services in rural areas \cite{park2022network}.
Recently, an efficient online service function chain deployment algorithm was proposed for dynamic network function virtualization in \ac{NIB}s for industrial applications \cite{sun2020dynamic}. Likewise, 6G-enabled \ac{NIB}'s channel characteristics were explored for the internet of connected vehicles to realize full-coverage, full-spectrum, and full applications \cite{lv20206g}. Another industrial application was the secure decentralized spatial crowd-sourcing for 6G-enabled \ac{NIB}s, which enabled the collection/transmission of security information on the blockchain using \ac{NIB}, without depending on the third party \cite{zhang2021secure}. Moreover, EmergeNet provided reliable and repidly deployable small-scale cellular network for emergency and disaster scenariosx which was based on self-organizing network to enable autonomous decision-making for optimal network functioning \cite{iland2014emergenet}. \ac{NIB}s were used for remote coverage and emergency deployment owing to their cost effective design and deployment \cite{heimerl2013local}. }

None of the existing works has studied the deployment issue of NIB. However, this is important for its efficient operation. In this paper, we jointly address the \ac{NIB} deployment, resource allocation, service provisioning, and backhauling challenges. In this context, multiple UAV-borne \ac{NIB} (UAVB-NIB) hover over the area of interest where each UAVB-NIB is equipped with downward facing antenna arrays for communication with GUs and upward facing antenna for connection with the HAPs. Highly directional antenna at UAVB-NIBs is used for backhaul link so that it can direct beam to HAPS mitigating the free-space path losses (FSPL). For backhaul, HAPS employs \ac{NOMA} to serve different UAVB-NIBs that share the same spectral and temporal resources. In addition, each UAVB-NIB employs \ac{RZF} precoding for \ac{GU}s accessing same \ac{RAT} in its coverage area. A holistic approach and suitable algorithms to tackle the aforementioned challenges while guaranteeing user's QoS, fairness, energy, and spectral efficiency are proposed. This work presents a novel sequential algorithm that can solve the problems of deployment, \ac{UA}, beam optimization and resource allocation to maximize the system performance with limited power, spectrum and time resources. The main contributions of this work are: 
\begin{itemize}
\item  {Deployment problem is solved using Geometric Disk Cover (GDC) algorithm which determines the minimum number of required UAVB-NIBs and their locations. Moreover, the \ac{UA} problem is resolved using greedy algorithm with the objective to maximize the SINR. Each user is served by one and only one spot beam for user fairness (UF) and interference mitigation.} 
\item  {After the \ac{UA}, UAVB-NIBs location optimization is conducted to minimize beam-width/beam-radius to serve the associated users. The reduced beamwidth renders directive beam with high power density and antenna gain.}
\item For backhaul, HAPS employs phased antenna arrays and NOMA scheme to serve all the UAVB-NIBs in its coverage area \cite{Nokiahsieh2020uav}. The closed-form solution to the NOMA power allocation problem is presented. 
\item For AL, we employ \ac{SCA} to iteratively allocate power to different RAT users with the target to maximize the achievable sum rate under QoS constraints and backhaul limitations. 
\item The achievable data rate, energy efficiency (EE), spectral efficiency (SE) and UF of the proposed system are investigated to quantify the performance gains of the proposed algorithms over the existing ones.
\end{itemize}
\begin{figure}[t]
    \centering
   \includegraphics[width=0.8\linewidth]{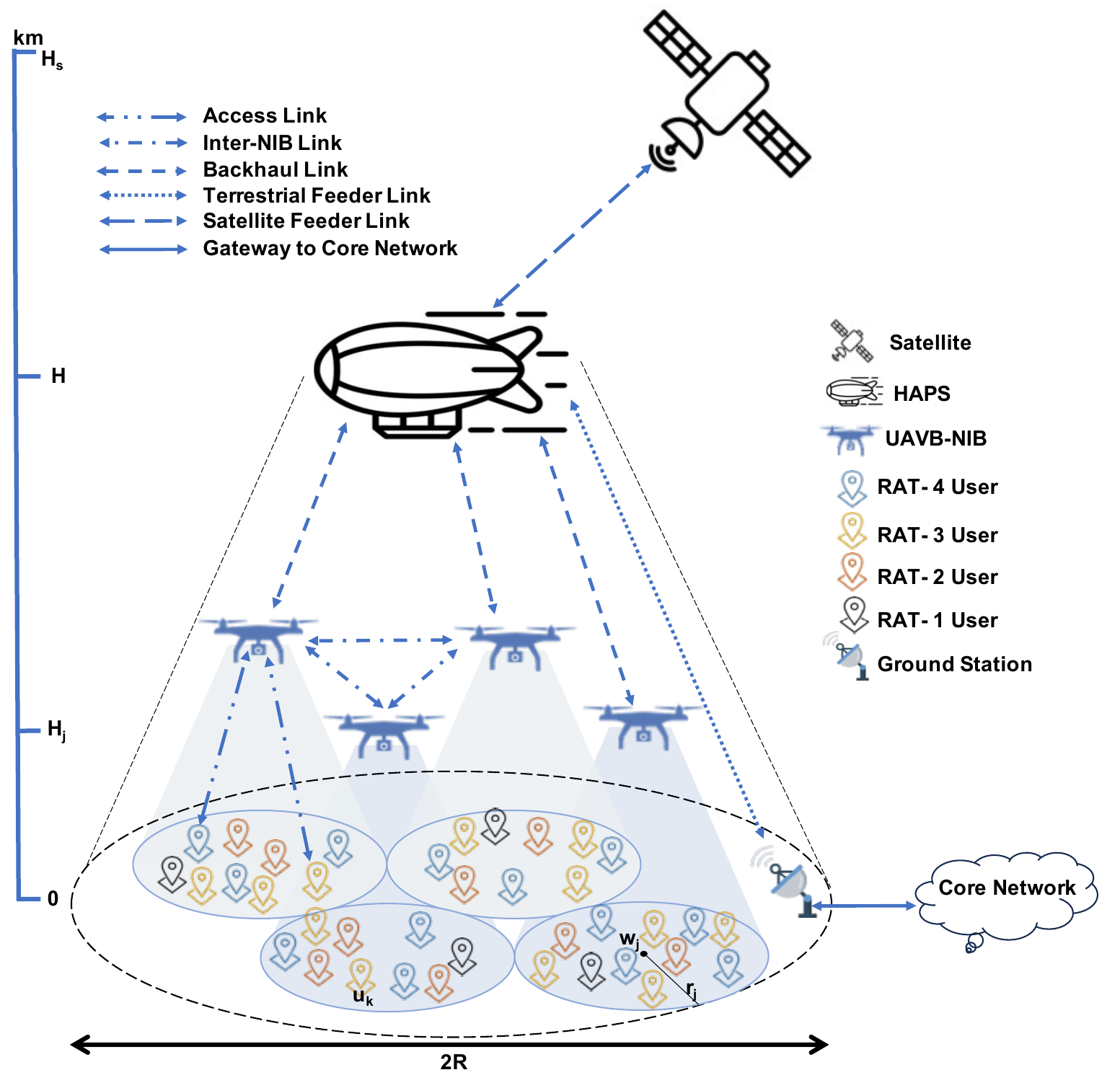}
    \caption{ \ac{NIB}-based Aerial Communication System  }
    \label{fig:HAPS}
\end{figure}
The rest of the paper is organized as follows: Section II describes the detailed system model for the communication between HAPS to UAVB-NIBS and between UAVB-NIBs and the GUs. Section III details the propagation model and link budget to incorporate both small-scale and large-scale fading effects in the access and backhaul channel. Next, achievable sum rates for the access and backhaul links are determined in Section IV. Subsequently, Section V formulates and solves the optimization problem. Numerical results are illustrated in Section VI followed by the conclusions and acknowledgements in Section~VII and VIII, respectively. 
\section{System Model}\label{SecII}
In order to offer aerial coverage to a remote circular area of radius $R$ having $K$ GUs with coordinates $u_k \in \mathbb{R}^2 \; \forall k$ (on the horizontal plane), \ac{NIB}s are mounted on UAVs for quick deployment. The coverage is offered by a fleet of $J$ UAVB-NIBs, each hovering at an altitude $H_j$ and covering a circular ground area with center ${\bf w}_j$ and 2D horizontal radius $r_j$  where $j \in \lbrace 1,2,\ldots,J \rbrace$, as shown in Fig \ref{fig:HAPS}. Each cell is served by a highly-directional and flexible beam from UAVB-NIB, using down-facing $M$ antennas, allowing frequency reuse in the neighboring cells for efficient spectrum allocation and minimal interference. We employ software-defined network based \ac{NIB} where each \ac{NIB} $j$ is capable of operating at various RATs $\Omega \in \lbrace \text{WiFi, 3G, 4G, 5G} \rbrace$ to serve $K_j^{\Omega}$ compatible users distributed uniformly in its coverage area. The number of users demanding RAT $\Omega$ follows the probability mass function $\Pr_{\Omega}$. We use the association parameters $\alpha_k^j \in \lbrace 0,1 \rbrace$ and $\beta_k^{\Omega} \in \lbrace 0,1 \rbrace$ to indicate that the user $k$ is associated with $j^{\rm th}$ UAVB-NIB operating at RAT $\Omega$, respectively, at a given time. We restrict $\sum_j \alpha_k^j = 1 \; \forall \; k$ for UF and $ \sum_{\Omega} \beta_k^{\Omega} = 1 \; \forall \; k$ for effective resource allocation. The received signal at user $k$ from $j^{\rm th}$ \ac{NIB} operating at RAT $\Omega$ is given~by:
\begin{align}\label{eq.LAPSy}
y_{kj}^{\Omega} = & \mathbf{h}_{kj}^{\Omega H} \mathbf{w}_{kj}^{\Omega}  \alpha_k^j \beta_k^{\Omega}  \sqrt{p_{kj}^{\Omega} P_{j}^{\Omega}} s_{kj}^{\Omega} +  \nonumber \\ 
& \sum_{l=1 \atop l \neq k}^{K}  \mathbf{h}_{kj}^{\Omega H} \mathbf{w}_{lj}^{\Omega}   \alpha_l^j \beta_l^{\Omega} \sqrt{p_{lj}^{\Omega} P_{j}^{\Omega}} s_{lj}^{\Omega}+  n_{kj}^{\Omega},
\end{align}
 where $\mathbf{h}_{kj}^{\Omega} \in \mathbb{C}^{M}$ is the channel vector between $M$ antennas of $j^{\rm th}$ \ac{NIB} and the single-antenna user $k$ operating at RAT $\Omega$, whereas $\mathbf{w}_{kj}^{\Omega} \in \mathbb{C}^{M}$ is the pre-coding vector for user $k$.   
 Moreover, $p_{kj}^{\Omega}\in \mathbb{R}$ and $s_{kj}^{\Omega}\sim \mathcal{CN}(0,1)$ are the allocated power coefficient and information bearing transmit signal for user $k$ from \ac{NIB} $j$ on RAT $\Omega$. Furthermore, $P_{j}^{\Omega}$ is the total transmission power budget of \ac{NIB} $j$ for RAT $\Omega$. 
One has $\sum_k \alpha_k^j \beta_k^{\Omega}  p_{kj}^{\Omega} = 1 $ for all users in $j^{\rm th}$ cell using RAT $\Omega$ to ensure the expenses are within the available power budget. 
In addition, the receiver thermal noise is modeled as a circular symmetric complex Gaussian random variable, i.e., $n_{kj}^{\Omega} \sim  \mathcal{CN}(0,\sigma_{kj}^{\Omega 2})$. Let's define $\mathbf{H}_j^{\Omega} =\left[ \mathbf{h}_{1j}^{\Omega},\mathbf{h}_{2j}^{\Omega}, \ldots, \mathbf{h}_{K_j j}^{\Omega} \right] \in \mathbb{C}^{M {\rm x} K_j}$ as the channel matrix comprising of channel coefficients between $M$ transmit antennas of  $j^{\rm th}$ UAVB-NIB operating at RAT $\Omega$ and 
the $K_j$ associated users whereas  $\mathbf{W}_j^{\Omega} =\left[ \mathbf{w}_{1j}^{\Omega},\mathbf{w}_{2j}^{\Omega}, \ldots, \mathbf{w}_{K_j j}^{\Omega} \right] \in \mathbb{C}^{M {\rm x} K_j}$ being the precoding matrix for all connected users, such that the $ \mathbf{w}_{kj}^{\Omega}$ intended for user $k$ is orthogonal to every channel vector $ \mathbf{h}_{lj}^{\Omega}$ associated with users $l \neq k$.
Then, the received signal vector can be written as ${\bf y} = \mathbf{H}_j^{\Omega H}  \mathbf{W}_j^{\Omega} {\bf x} + {\bf n}$, where ${\bf x} $ and $ {\bf n}$ are the input signal vector and thermal noise vector at the receiver, respectively, with entries $ \alpha_k^j \beta_k^{\Omega}  \sqrt{p_{kj}^{\Omega} P_{j}^{\Omega}} s_{kj}^{\Omega} $ and $n_{kj}^{\Omega}$ for all associated users with $j^{\rm th}$ UAVB-NIB which are operating at RAT $\Omega$. Therefore, the \ac{RZF} pre-coder can be determined as~\cite{peel2005vector}
\begin{equation}
\mathbf{W}_j^{\Omega}= \zeta_j^{\Omega} ( \mathbf{H}_j^{\Omega} \mathbf{H}_j^{\Omega H} + \omega \mathbf{I}_M )^{-1}  \mathbf{H}_j^{\Omega},
\end{equation}
 where $\omega$ is the regularization scalar and $\zeta_j$ is the normalization scalar satisfying 
 \begin{equation}
 \zeta_j^{\Omega 2}\!\! = \! \! \frac{1}{ \text{tr} [  ( \mathbf{H}_j^{\Omega}  \mathbf{H}_j^{\Omega H} \! +\!\omega \mathbf{I}_M )^{-1}  \mathbf{H}_j^{\Omega} \mathbf{H}_j^{\Omega H}  ( \mathbf{H}_j^{\Omega} \mathbf{H}_j^{\Omega H}\! +\! \omega \mathbf{I}_M )^{-1}     ]}.
 \end{equation}
It is important to note that user $k$ will only experience co-channel interference from  users operating at the same RAT located within the same UAVB-NIB coverage area/cell. 
There is no inter-RAT interference as they operate at different frequency bands. Thus, the SINR $\gamma_{kj}^{\Omega}$ of the received signal $y_{kj}^{\Omega}$ can be derived as 
\begin{equation}\label{SINR}
\gamma_{kj}^{\Omega} = \frac{ | \mathbf{h}_{kj}^{\Omega H} \mathbf{w}_{kj}^{\Omega}|^2 \alpha_k^j \beta_k^{\Omega}  p_{kj}^{\Omega}}  
 {\sum_{l=1 \atop l \neq k}^{K} | \mathbf{h}_{kj}^{\Omega H} \mathbf{w}_{lj}^{\Omega}|^2   \alpha_l^j \beta_l^{\Omega} {p_{lj}^{\Omega}}  +  (\bar \gamma_{kj}^{\Omega})^{-1}    
},
\end{equation}
where, $\bar \gamma_{kj}^{\Omega} = P_{j}^{\Omega}/\sigma_{kj}^{\Omega 2}$ is the transmit SNR with noise variance ${\sigma_{kj}^{\Omega 2}}({\rm dBm}) = -174 + 10\log(B_j^{\Omega}) + {\rm NF_k}$. Here, $B_j^{\Omega}$ depicts the allocated channel bandwidth by $j^{\rm th}$ UAVB-NIB for RAT $\Omega$ and ${\rm NF_k}$ denotes the noise figure of the $k^{\rm th}$ user \cite{shibata2020system}.
 \begin{figure}[t]
    \centering
   \includegraphics[width=0.9\linewidth]{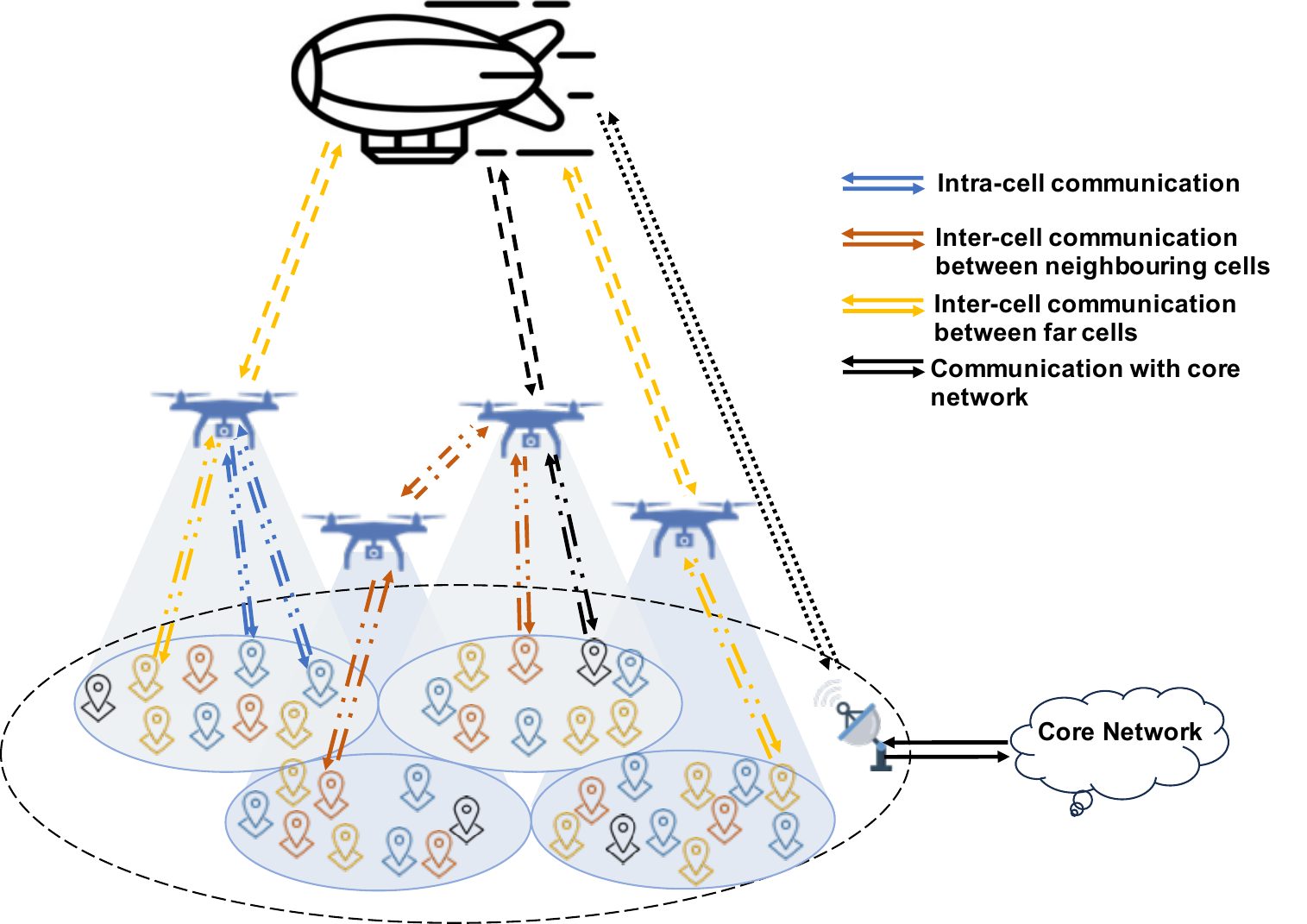}
    \caption{ Different Communication Scenarios  }
    \label{fig:CS}
\end{figure}
Fig. \ref{fig:CS} illustrates different communication scenarios in the AL:
Scenario I is the intra-cell communication within the UAVB-NIB coverage area, Scenario II is the inter-cell communication between GUs through the neighboring \ac{NIB}s, Scenario III is the inter-cell communication of GUs between distant UAVB-NIBs, and Scenario IV is the communication with core network. Interestingly, the \ac{NIB} base stations are capable of independent routing and communicating in Scenarios I and II, whereas in Scenarios III and IV the communication is carried through the backhaul link.

For the backhaul, consider a typical unmanned solar-powered quasi-stationary HAPS at an altitude $H$ over the desired coverage area with radius $R$ ranging between $60$km-$400$km. The HAPS operates from the stratospheric location (preferably between $18$km-$24$km), pertaining to the suitable atmospheric conditions for the stable flight operation. HAPS provides backhaul links to $J$ UAVB-NIBs over the fourth-generation ($4$G) long-term evolution (LTE) or $5$G new radio (NR) air interfaces.  {It can further be connected to the terrestrial or satellite gateway through the RF feeder link \cite{belmekki2024cellular} or free-space optical communication link \cite{ata2022haps}}. HAPS communication panel employs phased antenna arrays to fixate the coverage relative to the station-keeping flight pattern. Within the coverage area, the channel gain varies by increasing the distance from the beam center. The strongest channel gain is available along the boresight direction $\theta = 0$. However, as the distance varies and/or the azimuth direction deviates from the boresight, the performance can be degraded due to the increased path loss and reduced antenna radiation pattern gain. The striking difference in the channel gains of the connected UAVB-NIBs enables us to reap maximum benefits offered by non-orthogonal multiple access (NOMA).
Considering the DL-NOMA scenario, where the UAVB-NIBs are served by a directional beam with superposition coding as \footnote{The transmitted/received signals, channel gains and allocated powers are function of time. However, the time notation is omitted for brevity.}
\begin{equation}
\bar x = \sum_{j=1}^{J}{\sqrt{ f_j P_H} x_j},
\end{equation}
where $P_H $ is the available power budget for transmission after deducting the aerodynamics, electronics, and night-time operational expenses from the available solar power at a given time \cite{javed2023interdisciplinary}. 
Moreover, $f_j$ and  $x_j$ are the fraction of power allocated to and intended information signal for the $j^{\rm th}$ UAVB-NIB, respectively. It is important to highlight that $\sum_{j=1}^{J}{f_j} \leq 1$ in order to limit the power division within given budget. Thus, using conventional wireless communication model, the received signal at $j^{\rm th}$ \ac{NIB} from the HAPS is given by
\begin{equation}
v_{j} = g_j \sqrt{f_j P_{H}} x_j + g_j \sum_{i = 1 \atop i \neq j}^{J} \sqrt{f_i P_{H}} x_i + z_j,
\end{equation} 
where, $g_j$ is the channel gain coefficient between the HAPS array panel and $j^{\rm th}$ UAVB-NIB and $z_j$ is the receiver thermal noise modeled as circular symmetric complex Gaussian random variable, i.e., $z_j \sim  \mathcal{CN}(0,\sigma_j^2)$. Clearly, the $j^{\rm th}$ \ac{NIB}  receives the superposed signal and retrieves its own signal using the ordered arrangement $j_1, j_2, \ldots, j_J$ depending on their increasing channel strengths. 
UAVB-NIB performs SIC by first decoding the information from \ac{NIB} $1$ to ${j-1}$ and then subtracting it from the received signal. Thus, it can decode its own signal from the resultant by considering the interference from $j+1$ to $J$ as noise. Therefore, the signal-to-interference noise ratio $\Gamma_j$ for the HAPS-\ac{NIB} link at the $j^{\rm th}$ UAVB-NIB is given by
\begin{equation}\label{eq.SINR1}
\Gamma_j = \frac{|g_{j}|^2{{f_j P_H }}}{|g_{j}|^2 P_H \sum_{i= j+1}^{J}{{f_i }  } +\sigma_j^2},
\end{equation}
whereas, the \ac{NIB} with the strongest channel gain can successfully decode the information of all other UAVB-NIBs, rendering the SINR
\begin{equation}\label{eq.SINR2}
\Gamma_J = \frac{|g_{J}|^2{{f_J P_H }}}{ \sigma_J^2},
\end{equation}
where the noise power is given as
\begin{equation}
{\sigma_j^2}({\rm dBm}) = -174 + 10\log(B_H) + {\rm NF}_j,
\end{equation}
with $ {\rm NF}_j$ denoting the noise figure of the $j^{\rm th}$ \ac{NIB}  \cite{shibata2020system} and $B_H$ depicting the HAPS channel bandwidth of the backhaul link.   {
It is noteworthy that this work focuses on the downlink communications from HAPS to UAVB-NIBs and then from UAVB-NIBs to GUs. The uplink communication is reciprocal of the downlink communication, however, it will have an additional constraint to bound the transmit data-rate of the associated users according to the backhaul capacity limit, in order to avoid the delays at \ac{NIB}s. }
\section{Link Budget}
\label{SecIII}
In the aerial communication system, the radio signal propagation from aerial platform undergoes both small-scale and large-scale multipath fading. Moreover, the link channel gain is also dependent on the transmitter antenna gain and the receiver position. In this section, we will carry out  link budgeting to model the propagation loss for both HAPS-\ac{NIB} and \ac{NIB}-GU links.
\subsection{\ac{NIB}-GU AL}
The link between the UAVB-NIBs and GUs experiences FSPL and multipath fading due to the vertical distance between them and obstacles around the UE, respectively. Thus, each channel coefficient $h_{kj}^{\Omega}$ in the channel vector ${\bf h}_{k j}^{\Omega}$ can be expressed as:
\begin{equation}\label{eq.NU}
|h_{kj}^{\Omega}|^2 = \frac{ |{\tilde h}_{kj}^{\Omega}|^2 { G_{jk}^{\Omega}}}{{L_{jk}^{\Omega}}}, 
\end{equation}
where ${\tilde h}_{kj}^{\Omega}$ is the small-scale fading coefficient between the $j^{\rm th}$ UAVB-NIB and $k^{\rm th}$ user operating at RAT $\Omega$. The AL is assumed to be non-LOS dominant and hence $|{\tilde h}_{kj}^{\Omega}|$ is modeled as a Rayleigh distribution with scale parameter $\mho$ having independently and identically normal distributed real and imaginary components i.e., $\Re \lbrace {\tilde h}_{kj}^{\Omega} \rbrace \sim \mathcal{N}(0,\mho^2)$ and $\Im \lbrace {\tilde h}_{kj}^{\Omega} \rbrace \sim \mathcal{N}(0,\mho^2)$. The probability density function of the Rayleigh distribution is well-known to be
\begin{equation}\label{eq.Rayleigh}
f(x\mid \mho)={\frac  {x}{\mho ^{2}}}\exp \left({\frac  {-x^{2}}{\mho ^{2}}}\right), \quad x\geq 0
\end{equation}
Moreover, UAVB-NIB employs antenna arrays so as to generate a single high-gain dynamic beam and the beam gain from the $j^{\rm th}$ \ac{NIB} to user $k$ is given by $G_{jk}^{\Omega}$  which is mainly determined by the off-axis angle between the GU and the main lobe direction of the UAVB-NIB beam \cite{zheng2012generic}
\begin{equation}\label{eq.Gjk}
G_{jk}^{\Omega} = G_{\rm max} \left( \frac{\mathcal{J}_1 (\mu_{jk})}{2\mu_{jk}} + 36 \frac{\mathcal{J}_3 (\mu_{jk})}{(\mu_{jk})^3}    \right)^2,
\end{equation}
where $G_{\rm max}$ is the maximum beam gain along the boresight direction whereas $\mathcal{J}_1$ and $\mathcal{J}_3$ are the first-kind Bessel functions of order 1 and 3, respectively, and
\begin{equation}\label{eq.Mu}
\mu_{jk} = \frac{2.07123 \sin \theta_{jk}}{\sin \theta_j^{\rm 3dB}},
\end{equation}
with $\theta_j^{\rm 3dB}$ representing the one-sided half-power beam width of $j^{\rm th}$ UAVB-NIB transmit antenna and $\theta_{jk}$ marking the offset angle of user $k$ from the beam center. This allows on-demand dynamic beam as per the user distribution, which improves coverage efficiency.
In addition, $L_{jk}^{\Omega}$ in \eqref{eq.NU} is the path loss as a function of the distance between $j^{\rm th}$ UAVB-NIB and $k^{\rm th}$ user operating at RAT $\Omega$ as \cite{wang2020demand}: 
\begin{equation}
L_{jk}^{\Omega}[{\rm dB}] = \frac{A}{1+a\exp \left( {-b(\phi_{jk}-a)} \right)} + B_{jk}^{\Omega},
\end{equation}
where
\begin{align}
& A =  \eta_{\rm LOS} - \eta_{\rm NLOS}, \\
& B_{jk}^{\Omega} = 20 \log_{10}(d_{jk}) +20 \log_{10} \left( \frac{4 \pi f_c^{\Omega}}{c}  \right) +  \eta_{\rm NLOS}, \\
& \phi_{jk} =  \arcsin \left( \frac{H_j}{d_{jk}}  \right), 
\end{align}
with $\eta_{\rm LOS},\eta_{\rm NLOS}, a, \; {\text {and}} \; b$ denoting the constant environment-related parameters while $\phi_{jk}$, $c$, $f_c^{\Omega}$ and $d_{jk}$ denote the elevation angle of user $k$ from UAVB-NIB $j$, speed of light, carrier frequency of RAT ${\Omega}$ and distance between the user $k$ and UAVB-NIB $j$. Consequently, the linear pathloss in \eqref{eq.NU} is $L_{jk}^{\Omega} = 10^{L_{jk}^{\Omega}[{\rm dB}]/10}$. 
\subsection{HAPS-\ac{NIB} Backhaul}
The backhaul link also undergoes both small-scale fading ${\tilde g}_{j}$ and large-scale fading $L_j^{\rm HAPS}$
\footnote{Note that the HAPS station-keeping flight does not contribute to the fast fading since there is no moving scatter surrounding the aircraft \cite{Nokiahsieh2019propagation}. }. Hence, the channel coefficient $g_j$ can be expressed as follows:
\begin{equation}\label{eq.Link}
|g_j|^2 = \frac{ |{\tilde g}_{j}|^2 { G_j^{\rm HAPS}}}{{L_j^{\rm HAPS}}}, 
\end{equation}
where $|{\tilde g}_{j}|$ is assumed to be Ricean distributed with LOS dominant characteristics between HAPS and UAVB-NIBs. Its probability distribution is given as the magnitude of a circularly-symmetric non-central bivariate normal random variable such as \cite{cuevas2004channel,kanatas2017radio,oladipo2007stratospheric} 
\begin{equation}\label{eq.Rician}
f(x\mid \nu ,\sigma_{\rm f })={\frac  {x}{\sigma_{\rm f } ^{2}}}\exp \left({\frac  {-(x^{2}+\nu ^{2})}{2\sigma_{\rm f } ^{2}}}\right)I_{0}\left({\frac  {x\nu }{\sigma_{\rm f } ^{2}}}\right),
\end{equation}
where $I_0$ denotes the zeroth-order modified Bessel function of the first kind whose shape parameter $K_s$ is defined by the ratio between the average power of LOS component and the average power associated with NLOS multipath components i.e., $K_s = {{ {\nu ^{2}}/{2\sigma_{\rm f } ^{2}}}}$. The transmitter antenna gain $G_j^{\rm HAPS}$ from HAPS to the $j^{\rm th}$ UAVB-NIB depends on the antenna aperture efficiency $\eta$ , half-power beamwidth of the antenna $\theta_m^{\rm 3dB}$, HAPS altitude $H$, UAV altitude $H_j$ and the distance of the UAVB-NIB $j$ from the center of the HAPS beam $w_0$ with \cite{takahashi2019adaptive}
\begin{equation}\label{eq.Gain1}
\left[ G_j^{\rm HAPS} \right]_{\rm dB} = \left[ G_{\rm 0}^{\rm HAPS}  \right]_{\rm dB}  - 12 \frac{G_{\rm 0}^{\rm HAPS} }{\eta} \left(  \frac{\theta_{j}}{70 \pi} \right) ^2, 
\end{equation}
where the peak HAPS antenna beam gain is $G_{\rm 0}^{\rm HAPS} = \eta \left( 70 \pi /  \theta_{\rm HAPS}^{\rm 3dB} \right)^2 $ and the beam angle (angle of departure) of the $j^{\rm th}$ UAVB-NIB can be derived using
\begin{equation}\label{eq.Gain2}
\theta_{j} = \tan^{-1} \left( \frac{\|{ w}_j - w_0\|}{H-H_j}    \right).
\end{equation}
 Evidently, the antenna directivity gain reduces while moving away from the boresight position in a horizontal plane. Finally, the $L_j^{\rm HAPS}$ is the FSPL as a function of  the distance between HAPS and $j^{\rm th}$ UAVB-NIB  i.e., $d_l^m $. We employ the space communication model for the aerial HAPS to compute the received signal path loss $L_j^{\rm HAPS}$ as \cite{Space2017}
\begin{equation}
L_j^{\rm HAPS}= \frac{16\pi^2 d_j^2 }{\lambda^2},
\end{equation}
where $\lambda$ is the wavelength corresponding to the carrier frequency of HAPS. The FSPL renders the ratio between transmit power and received power.  
\section{Performance Measure}
The overall performance of the system depends on the capacity of the AL as well as the backhaul link. We can evaluate the system performance in terms of achievable sum rate in the access and backhaul link. 

We adopt a system where UAVB-NIB  serves all RAT $\Omega$ users in its coverage area simultaneously. They all share the same bandwidth $B_j^{\Omega}$ and transmit with different power and precoding vectors to employ RZF precoding at the receiver for error-free detection. Considering linear precoding, the information rate for user $u_{kj}^{\Omega}$ is given by
\begin{equation}
 R_{kj}^{\Omega} =   B_j^{\Omega}  \log_2 (1+\gamma_{kj}^{\Omega}).
\end{equation}
Consequently, the sum rate of the AL is given by $R_a = \sum_{j=1}^J \sum_{k=1}^K  \sum_{\Omega} R_{kj}^{\Omega}  $ which aggregates the downlink data rate from all the UAVB-NIBs to the associated users operating at the desired RAT technologies. Thus,
\begin{equation}
R_a =  \sum_{j=1}^J \sum_{\Omega} B_j^{\Omega} \sum_{k=1}^K \log_2 (1+\gamma_{kj}^{\Omega}).
\end{equation} 
On the other hand, the backhaul link employs DL-NOMA at the HAPS and each UAVB-NIB receives the superposed signal and performs SIC based on the channel strength  ordering to decode its own signal. Assuming perfect receiver \ac{CSI}, we get accurate \ac{NIB}-ordering and error-free decoding. Thus, the achievable rate of $j^{\rm th}$ UAVB-NIB is given by
\begin{equation}
R_j = B_H \log_2 \left[ 1+{\Gamma}_j  \right].
\end{equation}
conditioned on $R_{j \rightarrow l} > {\tilde R}_j \; \forall \; j \leq l$, where ${\tilde R}_j$ is the target data rate of the $j^{\rm th }$ \ac{NIB} while $R_{j \rightarrow l}$ denotes the rate of the $l^{\rm th }$ \ac{NIB} to detect $j^{\rm th }$ \ac{NIB}'s message when $j \leq l$ in \ac{NIB} ordering  i.e., 
\begin{equation}
R_{j \rightarrow l}  = B_H \log_2 \left( 1 +  \frac{|g_{l}|^2{{f_j P_H }}}{|g_{l}|^2 P_H \sum_{i= j+1}^{J}{{f_i }  } +\sigma_j^2} \right) \geq {\tilde R}_j. 
\end{equation}
Thus, the sum rate $R_b$ of all UAVB-NIBs can be written as $R_b = \sum_{j= 1}^{J}{ R_j}$ yielding
\begin{equation}
R_b = \sum_{j= 1}^{J}{ R_j} = B_H \sum_{j= 1}^{J} \log_2 \left[ 1+ {  \Gamma}_j   \right],
\end{equation}
where the received SINR at $j^{\rm th}$ \ac{NIB} in \eqref{eq.SINR1} can be expressed using \eqref{eq.Link} as
\begin{equation}\label{eq.SINR3}
\Gamma_j = \frac{{{f_j  }}}{\sum_{i= j+1}^{J}{{f_i }  } +\aleph_j },
\end{equation}
where, 
\begin{equation}
\aleph_j = \frac{\sigma_j^2 L_j^{\rm HAPS}}{P_H | \tilde{g}_{j}|^2 G_
j^{\rm HAPS}}.
\end{equation}
Likewise, the SINR of UAVB-NIB with the strongest channel gain $ {  \Gamma}_J $ in \eqref{eq.SINR2} can be manifested using \eqref{eq.Link} as
\begin{equation}\label{eq.SINR4}
\Gamma_J = f_J/\aleph_J. 
\end{equation}
\section{Problem Formulation and Solution}
This work aims to jointly optimize the sum rate of all users in the coverage area of HAPS while guaranteeing their quality-of-service (QoS), UF, and expenses within the available power budget.
This optimization problem is targeted at optimizing:  
\begin{enumerate}
\item UAV deployment: $J$ the number of UAVB-NIB to serve all users in the coverage area, the UAV locations ${\bf w}_j$ (i.e., ground projection or center of coverage cells in 2D horizontal plane) and favorable hovering altitudes $H_j$ of each UAVB-NIB.  
\item \ac{UA}: $\alpha_k^j$ the association variable between the users and the UAVB-NIBs for the given $\beta_k^{\Omega}$ i.e., the user demand for particular RAT  $\Omega$.
\item Beam optimization signifying  the $3$dB half-power beam-widths (HPBWs) ${\theta}_j$ and beam-radii $r_j$ of each participating UAVB-NIB.
\item NOMA power allocation factors $f_j$ for each UAVB-NIB. 
\item NIB power allocation factors $p_{kj}^{\Omega}$ for each user in its coverage area demanding a particular RAT service.  
\end{enumerate}
 We formulate the design problem to maximize the sum data rate of all users in the access downlink communication in the heterogeneous communication system as:
\begin{subequations}\label{eq.P1}
\begin{alignat}{2}
\textbf{P1}:\quad &\!\!\!\!\underset{J,{\bf w}_j,H_j,p_{kj}^{\Omega},f_j,\atop {\theta_j}, {r_j},{\alpha_k^j} \forall k,j}{\text{maximize }}
&& \sum_{j=1}^J  \sum_{k=1}^K  \sum_{\Omega} B_j^{\Omega} \log_2 (1+\gamma_{kj}^{\Omega})  \label{eq.Obj} \\
&\!\!\!\! \text{s.t.} 
& &  R_a (\gamma_{kj}^{\Omega}) \leq R_b (\Gamma_j), \; \forall \; k \in K_1,j,\Omega  \label{eq.AB} \\ 
& & & R_{{k}j}^{\Omega} \geq R_{\rm min},\quad \forall k,j, \Omega  \label{eq.Rth} \\ 
& & & \alpha_k^j \in  \lbrace 0,1\rbrace \&  \sum_{j=1}^{J} \alpha_k^j = 1, \quad \forall k,j  \label{eq.UA} \\
& & &  \sum_{j=1}^{J} \alpha_k^j \|{\bf u}_k - {\bf w}_j\| \!\! \leq \!\! \sum_{j=1}^{J}\!\! \alpha_k^j r_j,  \forall k  \label{eq.UA2} \\
& & &  \theta_{\rm min} \leq \theta_j^{3 {\rm dB}} \leq  \theta_{\rm max}, \forall j \label{eq.BW} \\
& & &  H_{\rm min} \leq H_j \leq  H_{\rm max}, \forall j \label{eq.NA} \\
& & & R \geq  r_j \geq 0.443 \lambda^{\Omega} H_j / D, \forall j  \label{eq.BR} \\
& & &    R_{j \rightarrow  l} \geq \tilde{R}_j, \, \forall j \leq l   \label{eq.SIC}   \\ 
& & &  {  \sum\nolimits_{k } \sum\nolimits_{\Omega}  \alpha_k^j \beta_k^{\Omega} p_{kj}^{\Omega} } \leq 1,  \quad \forall j \label{eq.UPF} \\
& &&  0 \leq{  p_{kj}^{\Omega}} \leq 1, \quad \forall k,j,\Omega  \label{eq.UPF2} \\
& & &    \sum \nolimits_{j }  f_j \leq 1,  \quad \forall j \label{eq.NPF} \\
& &&  0 \leq{  f_j } \leq 1, \quad \forall j  \label{eq.NPF2} \\
&&& f_1 \geq f_2 \geq \ldots \geq f_J   \label{eq.NO}  \\
&&& 1 \leq  J \leq K \label{eq.NN} 
\end{alignat}
\end{subequations}
The constraint \eqref{eq.AB} ensures that the sum rate of the GUs in the AL does not exceed the sum rate of the serving UAVB-NIBs in the backhaul link. The user set $K_1$ comprises of all users communicating in Scenarios III and IV in Fig. \ref{fig:CS}.
The QoS rate constraint \eqref{eq.Rth} ensures that each GU is guaranteed the minimum rate threshold $R_{\rm min}$. The constraint \eqref{eq.UA} restricts the Boolean entries $\alpha_k^j \in  \lbrace 0,1\rbrace, \forall k,j $ (where $1 \leq k \leq K$ and $1 \leq j \leq J$) and the \ac{UA} limits users to connect to any one UAVB-NIB at a given time for UF and effective utilization of the given resources. Moreover, the constraint \eqref{eq.UA2} ensures that the associating user resides within the beam coverage area of UAVB-NIB. The essential bounds on the 3dB HPBW, UAVB-NIB altitude, and beam radii,  are guaranteed by the constraints \eqref{eq.BW}, \eqref{eq.NA} and \eqref{eq.BR}, respectively. The spot beams of UAVB-NIB can be adjusted by the beamwidth control. It is worth noting that a narrower beam than the given bounds is not achievable with the given antenna array dimensions. In addition, the rate constraint \eqref{eq.SIC} warrants the successful information decoding of all NIBs with weaker channel gains at NIBs with strong channel conditions to avoid error propagation. Next, the constraints \eqref{eq.UPF} and \eqref{eq.UPF2} limit the total transmission power and individual power factor of each GU within the available power budget. The sum of power coefficients of all users operating at $\Omega$ within the same NIB coverage cannot exceed $1$. Likewise, the power limitations on NIB are given by \eqref{eq.NPF} and \eqref{eq.NPF2}. The optimal NIB power factor ordering in \eqref{eq.NO} allocates more power to the weak users and vice versa for UF. Lastly, the bounds on the number of deployed UAVB-NIB are constrained  \eqref{eq.NN}.  

The problem {\bf P1} is a non-convex mixed integer programming problem. Therefore, we divide this problem into sub-problems and solve these sub-problems sequentially and iteratively.
The subproblems are solved for fewer optimization parameters assuming that other design parameters are fixed.
\subsection{ UAVB-NIB Deployment }
The first challenge is to identify the number of UAVB-NIBs $J$ to cover a remote area of radius $R$ and user distribution $K$. Each NIB hovering at an altitude $H_j$ offers the coverage to the ground circular area using high-density narrow spot beam with beamwidth $\theta_j^{\rm 3dB} \in \mathbb{R} $, beam radius $r_j \in \mathbb{R}^+$, and beam location\footnote{Beam location indicates the center of the beam with maximum antenna gain in the boresight direction} ${\bf w}_j \in \mathbb{R}^2$ for all $j \in \{1,2,\ldots,J \}$ as detailed in Fig. \ref{fig:HAPS}. Interestingly, the variable $J$ ranges between $1 \leq J \leq K$, reflecting that there can be at least one NIB to serve all GUs or a maximum of $K$ UAVB-NIBs to individually serve each user. Both of these bounds are very loose. The lower bound is unrealistic for a large coverage area as the UAVB-NIB has a limited coverage whereas the upper bound is economically unsuitable. In essence, the value of $J$ is a trade-off between SINR and efficient allocation of resources. The higher value of $J$ renders highly directional beams with concentrated power, improved SINR and higher data rates but inefficient and uneconomical resource allocation. Therefore, we need to find the minimal number of UAVB-NIBs and their respective locations to cover all the users with a given user distribution.
The UAVB-NIB altitude and coverage radius are related to each other through half-power beamwidth of the spot beam as: 
\begin{equation}\label{eq.BR2}
r_j = H_j \tan \left( \frac{\theta_j^{\rm 3dB}}{2}  \right).
\end{equation}
Intuitively, for a given HPBW of NIB antennas, we can increase or decrease the UAV altitude to adjust the beam radius for suitable coverage. Our findings reveal that $R_a$  monotonically increases with $J$, but it is not economically feasible to deploy such a large number of UAVB-NIBs. The problem can have multiple solutions based on the economic constraints:

\begin{itemize}
\item Based on the relation $r_j = H_j \tan \theta_j$, we can use the lower bounds of the UAVB-NIB altitude and beam width to find the minimal possible beam radius.
\item Find $r_j \forall j$ which satisfies the constrained inequality \eqref{eq.AB} with equality i.e.,  $R_a = R_b$.
\end{itemize}
Problem {\bf P1} can be decomposed as the sub-problem \textbf{P1(a)} for optimal UAVB-NIBs deployment given the beam coverage radius $r_j \forall j$;
\begin{subequations}\label{eq.P1a}
\begin{alignat}{2}
\textbf{P1(a)}:\quad &\!\!\!\!\underset{J, {{\bf w}_j}}{\text{minimize }}
&&{J}  \label{eq.p20} \\
&\!\!\!\! \text{s.t.} 
& & \eqref{eq.UA},\eqref{eq.UA2},\eqref{eq.NN} 
\end{alignat}
\end{subequations}
This sub-problem finds the optimal locations (beam centers $w_m$) of the minimal number of beams required to cover the users in the disk of radius $R$ i.e., the coverage area of HAPS communication system. Problem {\bf P1(a)} is a well-known geometric disk cover (GDC) problem which aims to find minimum number of disks of given radius to cover a set of points in the plane. The famous GDC problem is NP-hard highlighting the NP-hardness of {\bf P1(a)}. 

The problem can alternately be reformulated as the identification problem for a set of points:
\begin{subequations}\label{eq.P1a1}
\begin{alignat}{2}
\textbf{P1(a1)}:\quad &\!\!\!\!\underset{{\bf z} \in \mathbb{B}^K}{\text{minimize} }
&&{ {\bf 1}^{\rm T} {\bf z}}  \label{eq.p20} \\
&\!\!\!\! \text{s.t.} 
& & {\bf z}_k \in  \lbrace 0,1\rbrace,  \forall k \\
&&& {\bf Dz} \geq {\bf 1} 
\end{alignat}
\end{subequations}
where ${\bf D}$ is the symmetric boolean matrix with entries
\begin{equation}\label{eq.GDC}
d_{kl} =
\begin{cases} 
1,  &  \|{\bf u}_k - {\bf u}_l\|_2 \leq r_j, \\
0,  & {\rm otherwise}. \\
   \end{cases}
\end{equation}
\begin{figure}[t]
  \centering
\subfigure[GDC Approach]{\label{fig1:WithGDC}\includegraphics[width=0.49\linewidth]{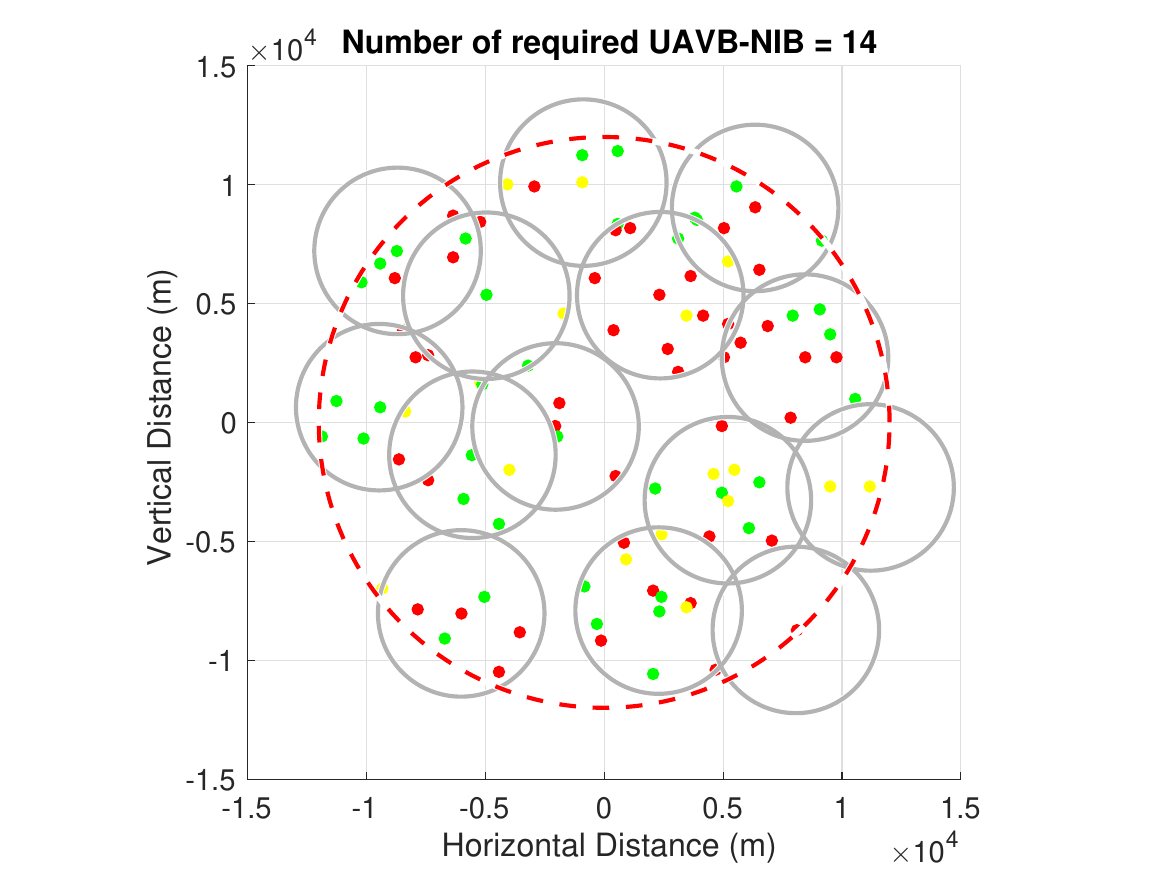}} 
\subfigure[Conventional Cellular Approach]{\label{fig1:WithoutGDC}\includegraphics[width=0.49\linewidth]{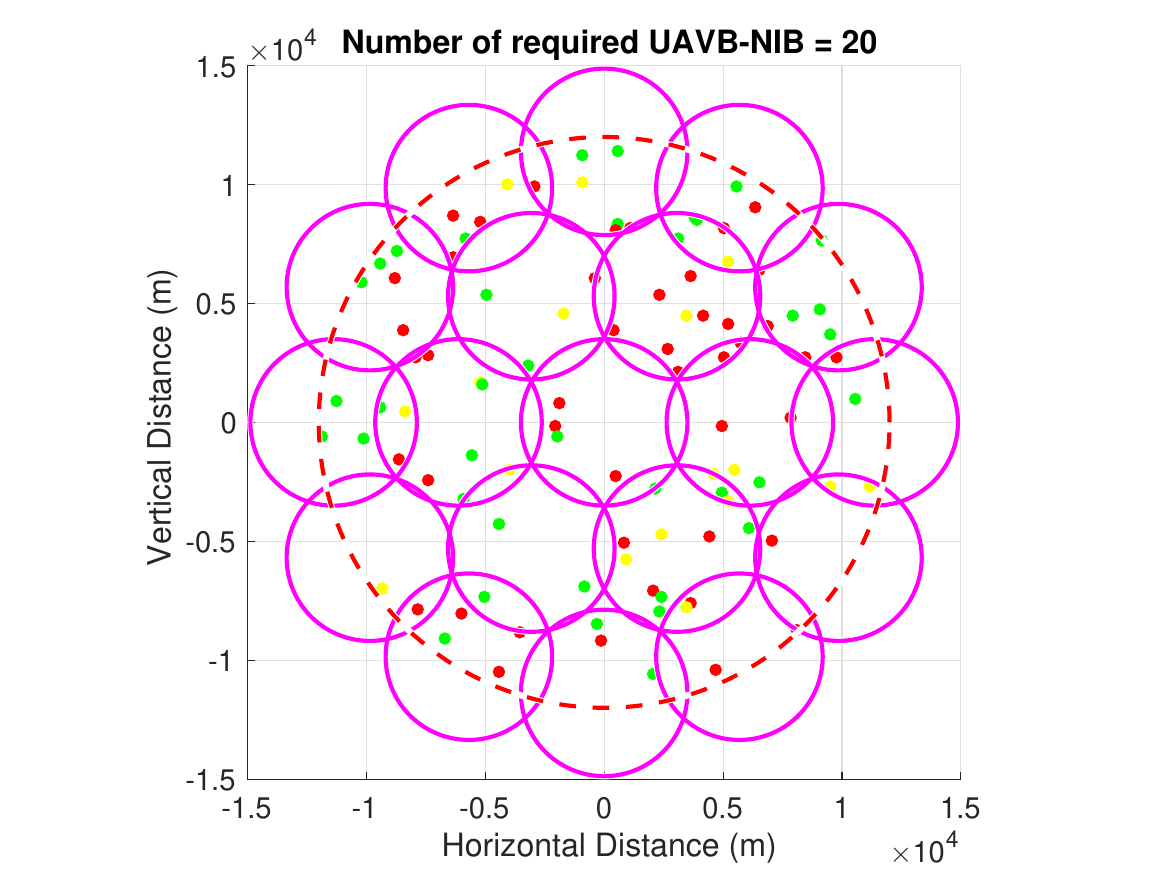}} 
\caption{Required number of UAVB-NIBs for user-connectivity in a given coverage area}
\label{fig:GDCImpact}
\end{figure}
To summarize, this problem identifies the minimum possible set of points from all the user coordinates such that if we draw the circles of radius $r_j$ around these points then it will encompass all the points in it's surrounding. We mark this set of points as the UAVB-NIB central positions and $\sum z^* = J^*$. This is a linear programming problem which can be easily solved using Lagrange function and KKT conditions for lower dimensions. The \textit{intlinprog} package in MATLAB can efficiently yield the UAVB-NIB deployment parameters for higher dimensional problems.  {Fig. \ref{fig:GDCImpact} illustrates an example of optimal deployment strategy for the coverage to a given set of users. GDC approach requires $14$ UAVB-NIBs in Fig. \ref{fig1:WithGDC} as opposed to the $20$ required UAVB-NIBs in Fig. \ref{fig1:WithGDC} with conventional cellular approach.} 
\subsection{Association Parameter}
We assume that all NIBs are capable of operating at all RATs depending on the demands from the associated users. In this context, any user can connect to any UAVB-NIB. However, for UF and higher system efficiency, each user is allowed to associate with one and only one UAVB-NIB.  Intuitively, the sum rate is maximized when the SINR $\gamma_{kj}^{\Omega}$ is maximized (for a given bandwidth for $\Omega$ RAT) owing to the monotonically increasing nature of logarithmic functions for $\gamma_{kj}^{\Omega}>0$. Hence, the \ac{UA} problem aims at finding the association parameters  $\alpha_k^j, \; \forall k,j$ can be restructured as subproblem \textbf{P1(b)}:
\begin{subequations}\label{eq.P1b}
\begin{alignat}{2}
\textbf{P1(b)}:\quad &\!\!\!\!\underset{\alpha_k^j, \; \forall k,j}{\text{maximize }}
&& {\sum_{j= 1}^{J}   \sum_{k = 1}^{K}  \sum_{\Omega }^{} \gamma_{kj}^{\Omega} }\left( {\alpha_{kj}^{\Omega}} \right)  \label{eq.p40} \\
&\!\!\!\! \text{s.t.} 
& &  \eqref{eq.UA} \; {\rm and} \; \eqref{eq.UA2}.
\end{alignat}
\end{subequations}
Constraint \eqref{eq.UA2} ensures that all the associated users of UAVB-NIB reside within its coverage area inside the main lobe of the antenna beam to be served simultaneously \cite{liu2020joint}. However, it is noteworthy that a user $k$ may reside within the radius of UAVB-NIB $j$ while being associated with another UAVB-NIB $j'$ in overlapping coverage zones. The solution to this association parameter problem can be quantified using the greedy algorithm as:
\begin{equation}
\alpha_k^j  = \begin{cases} 
 1, \qquad  \gamma_k^j \geq \gamma_k^{j'}  \; \forall j' \in \lbrace 1,2,\ldots,J \rbrace \backslash j &  \\
  0, \qquad  \text{otherwise}. & \\
   \end{cases}
\end{equation}
In practice, all the deployed UAVB-NIBs broadcast paging signals and the users can associate and connect to the one with the maximum received signal power \cite{alkhateeb2017initial}. 
\subsection{Location Optimization}
Consider the perfect \ac{CSI} and \ac{RZF} precoding to nullify co-channel interference between the users and decouple their information bearing signals. The UAVB-NIB localization problem is aimed at improving the channel gain between the UAVB-NIB and the associated users which will resultantly increase the SINR and AL rates in downlink communication scenario. The channel gain can be improved by adjusting the UAVB-NIB location parameters to enhance the scaling factors in \eqref{eq.NU} i.e., maximizing the antenna beam gain and minimizing FSPL. The higher antenna beam gain can be attained by using highly directional beams with concentrated radio frequency power, whereas, the FSPL can be decreased by adjusting UAVB-NIB altitude. Based on the derived $J$ and $\alpha_k^j$ for a given beam coverage radius $r_j \; \forall j$, we now carry out the location parameter optimization i.e., ${\bf w}_j, r_j, H_j, \theta_j^{\rm 3dB}$,and $\phi_j^{\rm 3dB}$ for the worst-case scenario i.e., maximizing the minimum channel gain and minimizing the maximum path loss. Interestingly, these variable are related rendering only three independent variables ${\bf w}_j, r_j,\phi_j^{\rm 3dB}$ while other location parameters can be derived from these independent parameters using $\theta_j^{\rm 3dB}= \pi/2 - \phi_j^{\rm 3dB}$ and $H_j = r_j  \tan (\phi_j^{\rm 3dB})$. 
\begin{subequations}\label{eq.P1c}
\begin{alignat}{3}
\textbf{P1(c)}:\quad &\!\!\!\!\underset{{{\bf w}_j, r_j},{\phi_j}^{\rm 3dB}}{\text{maximize }} \;&& \underset{k,\Omega} {\text{min}} \quad  \alpha_k^j ({ G_{jk}^{\Omega}[{\rm dB}] }- {L_{jk}^{\Omega}}[{\rm dB}] ) & \\
&\!\!\!\! \text{s.t.} 
&&   r_j \geq \text{max} \lbrace{\alpha_k^j \|{\bf u}_k - {\bf w}_j\|} \rbrace,  \forall k,j \label{eq.p31} \\ 
& & & \eqref{eq.BW}, \;  \eqref{eq.NA}, \; \eqref{eq.BR},\; {\rm and} \; \eqref{eq.BR2}
\end{alignat}
\end{subequations}
 The sub-problem \textbf{P1(c)} is disjoint problem for all $j$ UAVB-NIBs and hence it can be solved independently for all UAVB-NIBs with the given set of associated users. The joint optimization of \textbf{P1(c1)} is difficult and complex owing to the non-convex nature due to the mixture of bessel functions, exponential functions, sinusoidal functions and multiple optimization parameters. Thus, it is hard to achieve a global optimal solution. However, decomposing this problem into sub-problems is a promising choice to obtain sub-optimal solution close to the optimal one. We can then employ block coordinate descent (BCD) method for alternate optimization of the parameters to successively maximize the objective functions along one coordinate while fixing the local values at the other coordinates in each iteration. This method guarantees local stationary point because objective function is monotonically increasing in each coordinate with every iteration i.e., $f({\bf w}_j, r_j, {\phi_j}^{\rm 3dB}) \leq {\rm max} f({\bf w}_j, r_j| {\phi_j}^{\rm 3dB}) \leq f({\bf w}_j^*, r_j^*,  {\phi_j}^{\rm 3dB}) \leq {\rm max} f({\phi_j}^{\rm 3dB} |  {\bf w}_j^*, r_j^*  ) \leq f( {\bf w}_j^*, r_j^*, {\phi_j}^{{\rm 3dB}*})$.

Considering the objective function $f({\bf w}_j, r_j| {\phi_j}^{\rm 3dB})$ for $j^{\rm th}$ UAVB-NIB, the problem is equivalent to the minimum enclosing circle problem for a given set of points (users associated with $j^{\rm th}$ UAVB-NIB). Maximum sum rate can be achieved with maximum SNR by choosing minimum possible $r_j$ to enclose all the points in a circle centered at ${\bf w}_j$. This will ensure highly directional beam and concentrated power to serve the given set of users. This can be reformulated as a quadratic programming problem in \textbf{P1(c1)}
\begin{subequations}\label{eq.P1c1}
\begin{alignat}{3}
\textbf{P1(c1)}:\quad &\!\!\!\!\underset{{\bf \kappa} \in \mathbb{R}^{\varsigma_j}}{\text{minimize }}
&&  {\bf \kappa}^{\rm T} {\mathbf {\Xi}}_j {\bf \kappa} - { \bf d}_j^{\rm T} { \bf \kappa} &    \\
&\!\!\!\! \text{s.t.} &&  \sum_{n=1}^{\varsigma_j}  {\bf \kappa}_n = 1, \quad \kappa_n \geq 0  \quad \forall n 
\end{alignat}
\end{subequations}
where ${\bf \Xi}_j = {\bf U}_j{\bf U}_j^{\rm T}$ and ${ \bf d}_j = {\text {diag}}({\bf \Xi}_j )$ with 
\begin{align*}
{\bf U}= & [{\bf u}_1; {\bf u}_2; \ldots ;{\bf u}_K]  \in \mathbb{R}^{K {\rm x} 2}, \\
{\bf U}_j  \in \mathbb{R}^{\varsigma_j  {\rm x} 2}, & {\bf U}_j   \subseteq {\bf U},\lbrace {\bf u}_k \in {\bf U}_j | \alpha_k^{j}=1  \rbrace.
\end{align*}
 Moreover, $\varsigma_j = \sum_{k=1}^K \alpha_k^j$ is the number of users associated with $j^{\rm th}$ UAVB-NIB. The Lagrange function of the revised problem \textbf{P1(c1)} can be written as:
\begin{equation}\label{eq.LO1}
\mathcal{L}(\iota, \kappa ) = {\bf \kappa}^{\rm T} {\mathbf {\Xi}}_j {\bf \kappa} - { \bf d}_j^{\rm T} { \bf \kappa} - \iota ( \sum_{n=1}^{\varsigma_j}  {\bf \kappa}_n = 1);  \iota\geq 0.
\end{equation}
Solving this convex dual problem renders the primal and dual variables, which can then yield ${\bf w}_j^* = {\bf U}_j^{\rm T} \kappa^*$ and $r_j^* = \sqrt{ {\bf \kappa}^{*\rm T} {\mathbf {\Xi}}_j {\bf \kappa}^* - \iota^*}$. Based on the UAVB-NIB horizontal locations, we can now adjust individual altitudes using the optimal elevation angles $\phi_{j}^*\; \forall j$, by solving the following (see Appendix A) \cite{al2014optimal}
\begin{equation}\label{eq.LO2}
 \frac{A}{1\!\! +\! \!{\bar a} e^{ {-b \phi_{jk}^{\Omega}}}} + 20 \log_{10} (r_j \sec \phi_{jk}^{\Omega} )+  \bar{B}^{\Omega} = 0,
\end{equation}
where ${\bar a} = a e^{ab}$. Interestingly,  the evaluation of the optimal HPBW $\theta_{j}^{*\rm 3dB}$ and altitude $H_j^*$ for each participating UAVB-NIB is a disjoint problem and can be solved independently. Once the beamwidths are adjusted, the corresponding antenna beam gain for $k^{\rm th}$ user from the $j^{\rm th}$ UAVB-NIB can be evaluated using \eqref{eq.Gjk}. 
\subsection{Resource Allocation}
After UAV deployment, \ac{UA}, and beam optimization,
we focus on the design of power allocation parameters for the backhaul link. HAPS employs downlink NOMA strategy to simultaneously serve all the NIBs in its coverage area based on their locations and channel strength ordering. Thus, the power allocation problem for HAPS-NIB backhaul link can be written as: 
\begin{subequations}\label{eq.P1d}
\begin{alignat}{2}
\textbf{P1(d)}:\quad &\!\!\!\!\underset{\boldsymbol{f_j}}{\text{maximize }}
&&{\sum_{j= 1}^{J}   R_j }\left( \Gamma_j \right) \label{eq.50} \\
&\!\!\!\! \text{s.t.}  
& &    R_{j \rightarrow  l} \geq \tilde{R}_j, \, \forall j \leq l   \label{eq.51}   \\ 
& & &    \sum \nolimits_{j }  f_j \leq 1,  \quad \forall j \label{eq.52} \\
& &&  0 \leq{  f_j } \leq 1, \quad \forall j  \label{eq.53} \\
&&& f_1 \geq f_2 \geq \ldots \geq f_J   \label{eq.54}  
\end{alignat}
\end{subequations}
Given the UAVB-NIB ordering $N_1 \leq N_2 \leq \ldots \leq N_J $ with respect to their channel strengths, the target threshold constraint $ R_{j \rightarrow  l} \geq \tilde{R}_j, \, \forall j \leq l $ is most difficult to meet in the worst case scenario i.e., $j=l$. Thus, focusing on the worst case scenario, the constraint  \eqref{eq.51} can simplified as $R_{j} \geq \tilde{R}_j, \, \forall j$. Moreover, considering the same target threshold for all NIBs i.e., $\tilde{R}_j = R_{\rm th}$, we can present the closed-form solution to this problem as \cite{wang2019user}: 

For the given UAVB-NIBs, there exists a UAVB-NIB $\jmath$ in $1 \leq \jmath \leq J$, which satisfies the following condition:
\begin{equation}
\begin{cases} 
\left(  2^ {R_{\rm th}/B_H}-1 \right)   \left(\sum\limits_{i= \jmath}^{J}   \aleph_i 2^{(i-1)R_{\rm th}/B_H}   \right)  \leq 1, &  \\
\left(  2^{R_{\rm th}/B_H} -1 \right)   \left(\sum\limits_{i= \jmath-1}^{J}  \aleph_i  2^{(i-1)R_{\rm th}/B_H}  \right) \geq 1. & \\
   \end{cases}
\end{equation}
which indicates that the NIBs $\jmath$ to $J$ can achieve the target rate threshold $R_{\rm th}$ owing to the better channel conditions, whereas, the NIB $\jmath-1$ and weaker NIBs are unable to achieve the minimum threshold with the given power budget. Thus, we allocate the remaining power to the strongest NIB in order to maximize the sum data rate. 
 Hence, the maximum achievable sum rate ${R}_b^*$, the power coefficients of NIBs $f_j$ and the remaining power fraction $\Delta f $ are given by
\begin{equation}\label{eq.OptSR}
{R}_b^*  = \left(J-\jmath \right) R_{\rm th} + B_H \log_2 \left[ 1+  \frac{\Delta f}{    1-\Delta f +    \aleph_J   } \right],   
\end{equation}
\begin{equation}\label{eq.PowerAlloc}
\hat{f}_j = \left(2^{R_{\rm th}/B_H}-1 \right) \left( \sum\limits_{k= j+1}^{J}{\hat{f}_k} +   \aleph_j   \right),
\end{equation}
\begin{equation}\label{eq.Deltaf}
\Delta f = 1-\left(  2^ {R_{\rm th}/B_H} -1 \right)   \left(  \sum\limits_{i = \jmath }^{J} \aleph_i 2^{(i-1){R_{\rm th}/B_H}}  \right).
\end{equation}
The optimal sum rate in \eqref{eq.OptSR} comprises of two terms; the first term is the target rate threshold of NIBs $\jmath$ to $J$, whereas the second term is the additional rate of NIB $J$ using the leftover power $\Delta f$. This indicates that NIBs $\jmath$ to $J$ can attain target rates with assigned power coefficients $\hat{f}_{\jmath}$, $\hat{f}_{\jmath+1}$,..., $\hat{f}_J$ using \eqref{eq.PowerAlloc}. The remaining power with power allocation coefficient $\Delta f$ is insufficient for any of the remaining NIBs $1$ to $\jmath-1$ to fulfill their target threshold rate. Therefore, $\Delta f$ in \eqref{eq.Deltaf} is assigned to the strongest NIB $J$ in order to maximize the overall sum rate. 

Next is the resource allocation for the AL. The sum rate of the AL in scenarios III and IV is upper bounded by the rate of the backhaul link \eqref{eq.OptSR} derived by solving problem \textbf{P1(d)}. This subproblem is enumerated in problem \textbf{P1(e)}:
\begin{subequations}\label{eq.P1e}
\begin{alignat}{2}
\textbf{P1(e)}:\quad &\!\!\!\!\underset{p_{kj}^{\Omega} \forall k}{\text{maximize }}
&& \sum_{j=1}^J  \sum_{k=1}^K  \sum_{\Omega} B_j^{\Omega} \log_2 (1+\gamma_{kj}^{\Omega})  \label{eq.60} \\
&\!\!\!\! \text{s.t.} 
& & \tilde{R}_a (\gamma_{{k}j}^{\Omega}) \leq R_b^*, \quad \forall \; {k \in K_1},j,\Omega  \label{eq.61} \\ 
& & & {R}_{{k}j}^{\Omega} \geq R_{\rm min}, \quad \forall \; {k},j,\Omega  \label{eq.61b} \\ 
& & &  {  \sum\nolimits_{k } \sum\nolimits_{\Omega}  \alpha_k^j \beta_k^{\Omega} p_{kj}^{\Omega} } \leq 1,  \quad \forall j \label{eq.62} \\
& &&  0 \leq{  p_{kj}^{\Omega}} \leq 1, \quad \forall k,j,\Omega  \label{eq.63} 
\end{alignat}
\end{subequations}
where $\tilde{R}_a (\gamma_{kj}^{\Omega})$ is the subset of the AL rate comprising of the data rates of the set of users ${k} \in {K_1}$ in scenarios III and IV, which utilize the backhaul link for end-to-end communications. Assuming perfect \ac{RZF}, we get $| \mathbf{h}_{kj}^{\Omega H} \mathbf{w}_{lj}^{\Omega}|^2 = 0, \; \forall k \neq l $. Thus, using \eqref{SINR}, the objective function reduces to 
\begin{equation}
R_a (p_{kj}^{\Omega}) =  \sum_{j=1}^J  \sum_{k=1}^K  \sum_{\Omega} B_j^{\Omega} \log_2  \left( 1+ \bar \gamma_{kj}^{\Omega} | \mathbf{h}_{kj}^{\Omega H} \mathbf{w}_{kj}^{\Omega}|^2 \alpha_k^j \beta_k^{\Omega}  p_{kj}^{\Omega}\right).
\end{equation}
The simplified expression shows an ideal concave objective function as the weighted sum of the logarithmic functions and convex constraints without \eqref{eq.61}. For this non-convex constraint \eqref{eq.61}, we propose to employ successive convex approximation (SCA) and find the convex approximation $\tilde{R}_a$ and solve this problem iteratively. The first-order Taylor series approximation\footnote{s
First order Taylor series expansion of a function $f \left( x \right)$ around a point $ x^{(k)}$ is given as
\begin{equation}\label{eq22}
\tilde f\left( {x,x^{(k)} } \right) \approx f\left( {{{x}^{\left( k \right)}}} \right) + \nabla_x f\left( {{{x}^{\left( k \right)}}} \right)\left( {{x} - {{x}^{\left( k \right)}}} \right).
\end{equation}} $\tilde{R}_a$ around the power coefficient variables is given by:
\begin{equation}
\tilde{R}_a ({\bf p},{\bf p}^{(i)}) \approx \tilde{R}_a ({\bf p}^{(i)}) + \nabla_{\bf{p}} \tilde{R}_a({\bf p}^{(i)})^{\rm T} ({\bf p}-{\bf p}^{(i)}),
\end{equation}
where, ${\bf p}$ is the vector comprising of the power coefficients of all users depending on their UAB-NIB association and demanded RAT and ${\bf p}^{(i)}$ is the chosen/updated power coefficients vector at instant $(i)$. The gradient $\nabla_{{\bf p}}   \tilde{R}_a$ can be evaluated using the following partial derivatives:
\begin{equation}
\nabla_{\bf p}  \tilde{R}_a = \left[ \alpha_1^j \beta_1^{\Omega}\frac{{\partial  \tilde{R}_a }}{{\partial p_{1j}^{\Omega}}} \quad  \alpha_2^j \beta_2^{\Omega}\frac{{\partial  \tilde{R}_a}}{\partial {p_{2j}^{\Omega}}} \quad \ldots \quad  \alpha_K^j \beta_K^{\Omega}\frac{{\partial  \tilde{R}_a }}{{\partial p_{Kj}^{\Omega}}}  \right],
\end{equation}
where
\begin{equation}\label{eq23}
\frac {\partial  \tilde{R}_a}{  \partial   {p_{kj}^{\Omega}}} = \sum_j  \sum_{ \tilde{k}} \sum_{\Omega} \frac{ B_j^{\Omega}}{\log(2)}  \frac{     \bar \gamma_{kj}^{\Omega} | \mathbf{h}_{kj}^{\Omega H} \mathbf{w}_{kj}^{\Omega}|^2 \alpha_k^j \beta_k^{\Omega}  }{  1+ \bar \gamma_{kj}^{\Omega} | \mathbf{h}_{kj}^{\Omega H} \mathbf{w}_{kj}^{\Omega}|^2 \alpha_k^j \beta_k^{\Omega}  p_{kj}^{\Omega}}. 
\end{equation}
It is interesting to note that the gradient $\nabla_{p_{kj}^{\Omega}} \tilde{R}_a$ can be conveniently computed owing to the disjoint data rate of each user as a function of the allocated power fraction. This leads to the convex approximation of the constraint \eqref{eq.61}. Thus, problem $\textbf{P1(e)}$ can be iteratively solved using the successive convex approximation method. 
\begin{algorithm}[!t]
\caption{Sequential Optimization Algorithm}\label{Algo1}
\begin{algorithmic}[1]
\State \textbf{Input}: The number $\lbrace K \rbrace$, the coordinates of users $\lbrace {\bf u}_k \rbrace$ in the horizontal plane, radius of desired coverage area $\lbrace R \rbrace$, UAVB-NIB operable RATs $\Omega$, users RAT preference $\beta_k^{\Omega}$, NIB transmission power for each RAT $P_j^{\Omega} \;\forall \Omega$, HAPS transmission power $P_H$, UAVB-NIBs transmission power $P_j^{\Omega}$  and station-keeping altitude $\lbrace H \rbrace$.
\State \textbf{Initialize} 
 $i \gets 0$,  ${R}_a[i-1] \gets R_0$ and  $\epsilon \gets \infty$
 \State \textbf{Select} QoS minimum rate threshold $R_{\rm min}$, minimum possible beam radius $r_{\rm min}$, and regularization scalar $\omega$ 
  \State \textbf {Set} tolerance $\delta$,  $r[i] = r_{\rm min}$, and $r_{\rm UB} = R $
  \State \textbf {Choose} $\Delta r$ and identical beam radius $r_j[i] = r[i] \forall m$
\While {$\epsilon \ge \delta$ \& $r_{\rm min} \leq r_j[i] \leq R$} 
\State \textbf{Let}  ${i \gets i+1}$ 
\State \textbf{Update} $r_j[i]= r_j[i-1] + \Delta r$ for all UAVB-NIBs ensuring sequential increment with every iteration.    
\State \textbf{Determine} $M[i]$ and ${\bf w}_j[i] \; \forall \; m \in \left[  1,M \right]$ using GDC to solve \textbf{P1(a1)} in \eqref{eq.P1a1}  given constant $r[i]$. 
\State \textbf{Associate} users by solving \textbf{P1(b)} in \eqref{eq.P1b} to evaluate $\alpha_k^j[i]$ using greedy algorithm.
\State \textbf{Optimize} individual UAVB-NIB to valuate beam parameters $\tilde{w}_j[i]$, $\tilde{\theta}_j[i]$ and $\tilde{r}_j[i]$ by solving \textbf{P1(c1)} in \eqref{eq.P1c1}.
\State \textbf{Update}  ${w}_j[i] \gets \tilde{w}_j[i]$, $r_j[i] \gets \tilde{r}_j[i]$ and ${\theta}_j[i] \gets \tilde{\theta}_j[i]$. 
\State \textbf{Obtain} the available transmit power $P_H [i]$ of a solar powered HAPS at the chosen location on a given date and time of the day using the power estimation algorithms \cite{javed2023interdisciplinary}.
\State \textbf{Calculate} the channel coefficients $g_j[i]$ and $h_{kj}^{\Omega}[i]$ using the pathlosses $L_j^{\rm HAPS}[i]$ and beam gain $G_j^{\rm HAPS}[i]\; \forall j$ for the backhaul and pathloss   $L_{jk}^{\Omega}[i]$ and beam gain $G_{jk}^{\Omega}[i]\; \forall j,k,\Omega$ for the AL, respectively. 
\State \textbf{Compute} the power allocation coefficients  $f_j[i]$ for each UAVB-NIB using the closed form solutions of \textbf{P1(d)} in \eqref{eq.P1d} for the given QoS threshold $\tilde{R}_j$ and NIB ordering.
\State \textbf{Compute} the power allocation coefficients  $p_{kj}^{\Omega}[i]$ for each user by solving \textbf{P1(e)} in \eqref{eq.P1e}. 
\State \textbf{Evaluate} the sum rate in the AL $R_a[i]$ and backhaul sum rate $R_b[i]$
\State \textbf{Compare} ${R}_a[i]$ with $R_a[i-1]$
\State if ${R}_a[i] \geq R_a[i-1]$: $R_a^*[i] \gets {R}_a[i]$
\State if ${R}_a[i] \leq R_a[i-1]$: ${R}_a*[i] \gets R_a[i-1]$
\State Update  ${\epsilon \gets  {R}_a[i]- {R}_a[i-1]}$ 
\EndWhile
\State User grouping parameters: $J^* \gets J[i]$, ${\bf w}_j^* \gets {\bf w}_j[i]$
\State \ac{UA} parameters: $\alpha_{k}^{j*} \gets \alpha_{k}^{j*}[i] \; \forall \; l,m $
\State Beam radii: $r_j^* \gets r_j[i]$ 
\State Half-power beam widths: $\theta_j^* \gets \theta_j[i] \; \forall \;m$
\State Backhaul power allocation parameters: $f_j^{*} \gets f_j[i] \; \forall \; j$ 
\State AL power allocation parameters: $p_{kj}^{\Omega*} \gets p_{kj}^{\Omega}[i] \; \forall \; k,j,\Omega$ 
\State  Sum rate of GUs: $R_a^* \gets R_a[i]$
\end{algorithmic}
\end{algorithm}
The non-convex mixed integer programming problem $\textbf{P1}$ can be solved using the convexification of each individual sub-problem. Each sub-problem is then independently solved for fewer variables, assuming  the rest as given constants, as detailed in Algorithm \ref{Algo1}.  { Although the sub-problems are solved independently, however, the presented sequential order is crucial. It is pivotal to evaluate the serving number and locations of UAVB-NIB before user association. 
Deployment enables users to associate with the nearest UAVB-NIB rendering maximum signal strength. Likewise, UA helps in location optimization to form directive beams. Eventually, deployment and association permits the optimal resource allocation from the identified UAVB-NIB to the group of associated users demanding particular RATs.} 

\begin{figure}[t]
    \centering
   \includegraphics[width= 0.7\linewidth]{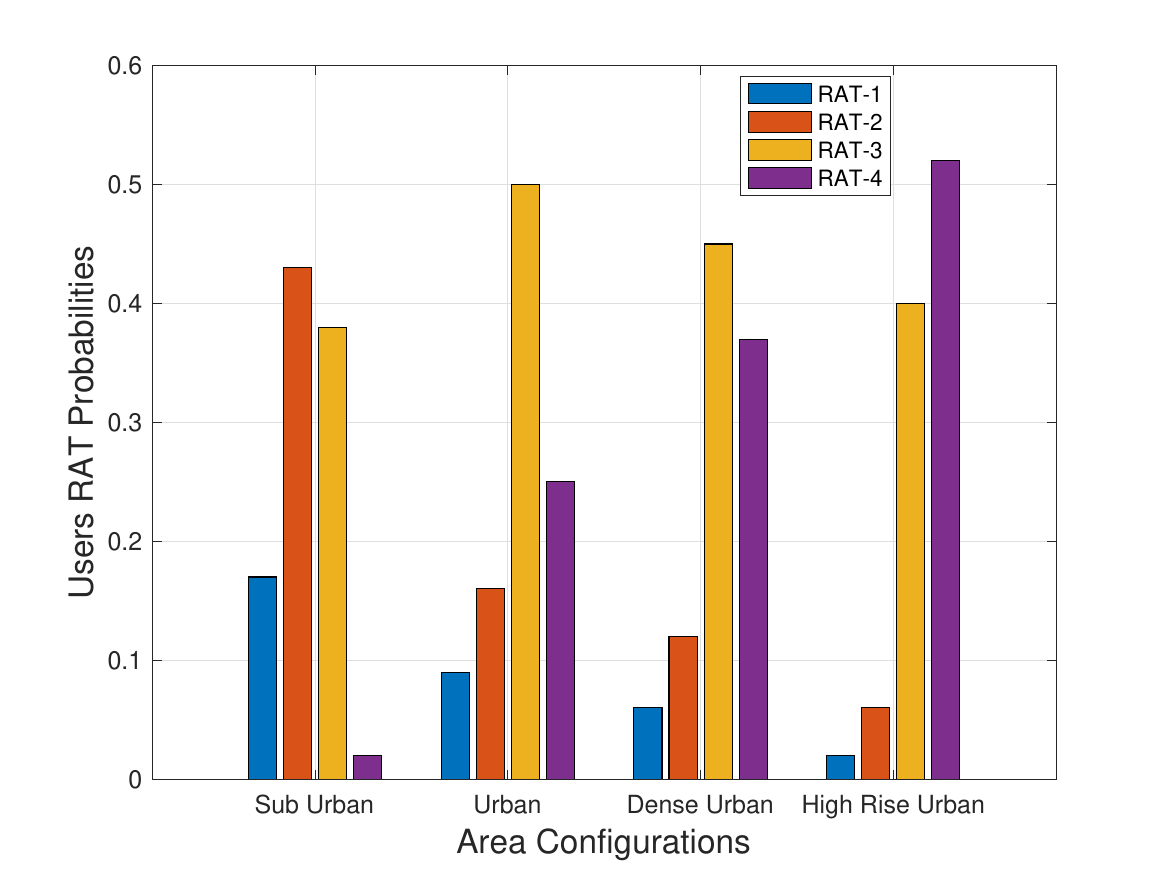}
    \caption{Users RAT Probabilities ($\Pr_{\Omega}$)}\vspace{-0.6cm} \label{fig:RAT}
\end{figure} 
 \begin{figure*}[htp]
  \centering
\subfigure[Users Distribution: Demanded-RAT]{\label{fig1:Users}\includegraphics[width=0.34\linewidth]{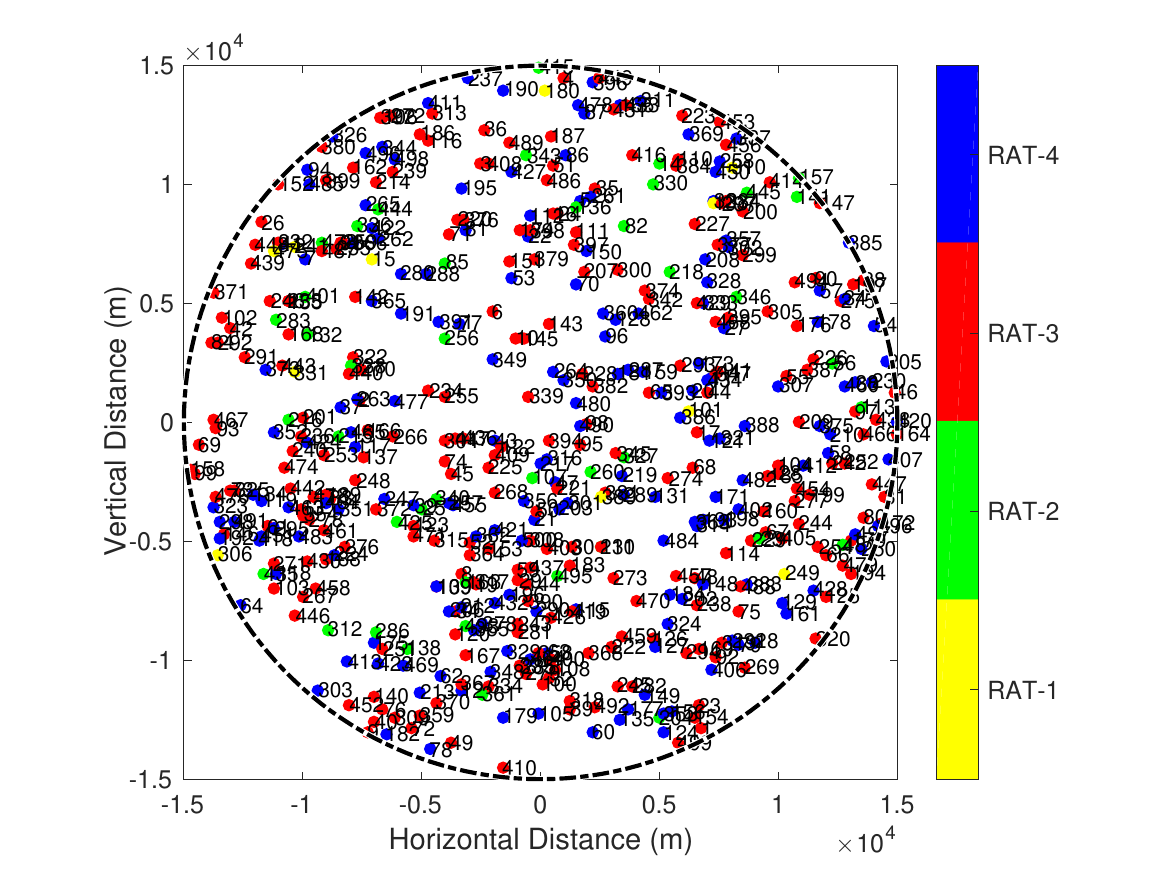}} \hspace{-0.5cm}
\subfigure[UAVB-NIBs: Deployment and Initial Coverage Zones]{\label{fig1:GDC}\includegraphics[width=0.34\linewidth]{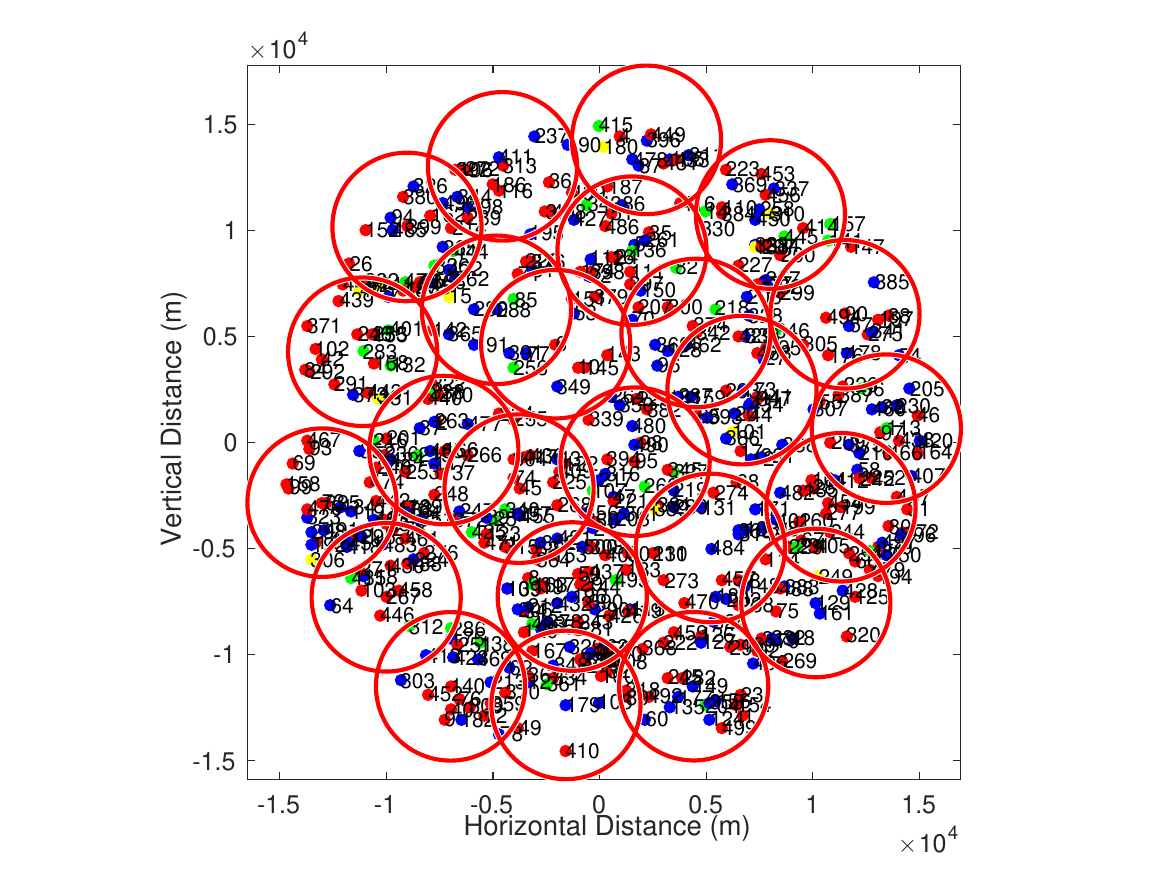}} \hspace{-1cm}
\subfigure[UAVB-NIBs: User Association]{\label{fig1:UA}\includegraphics[width = 0.34\linewidth]{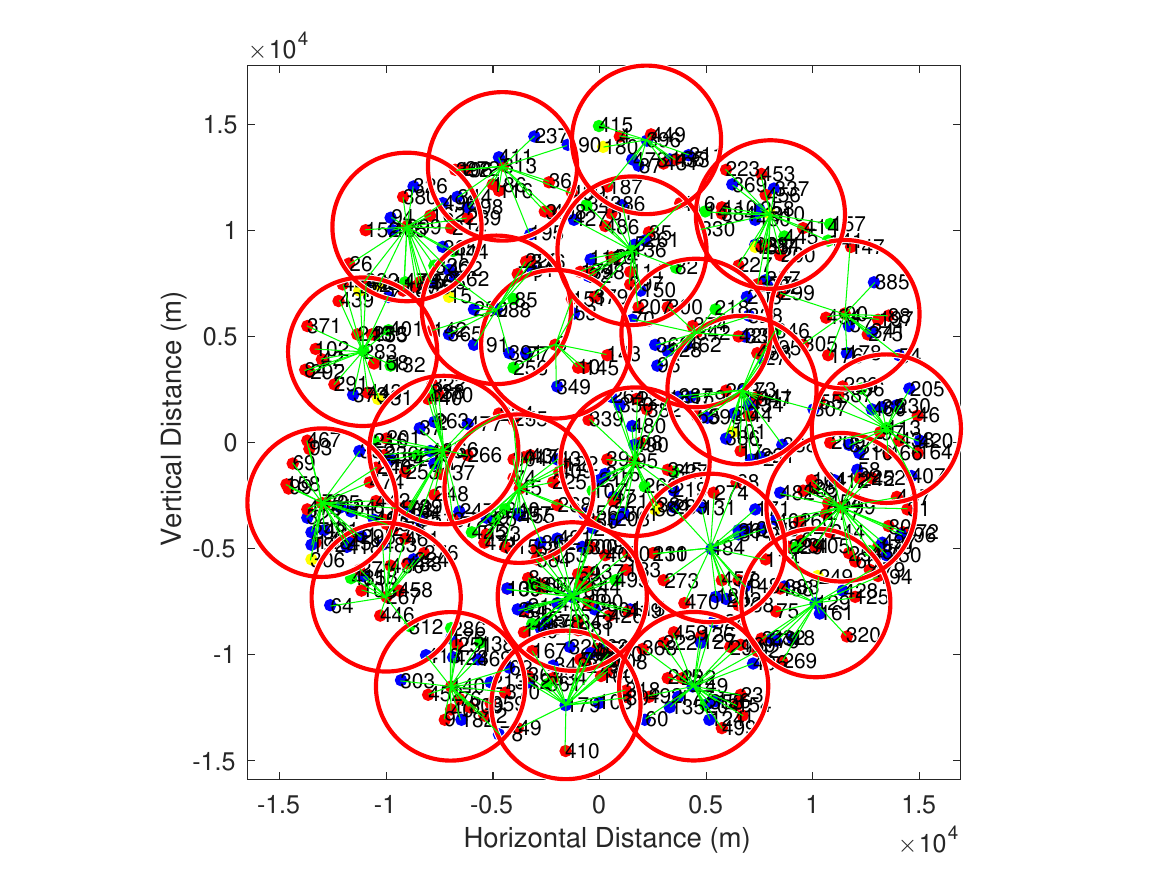}} \\
\subfigure[UAVB-NIBs: Location and Beam Optimization ]{\label{fig:BO}\includegraphics[width=0.36\linewidth]{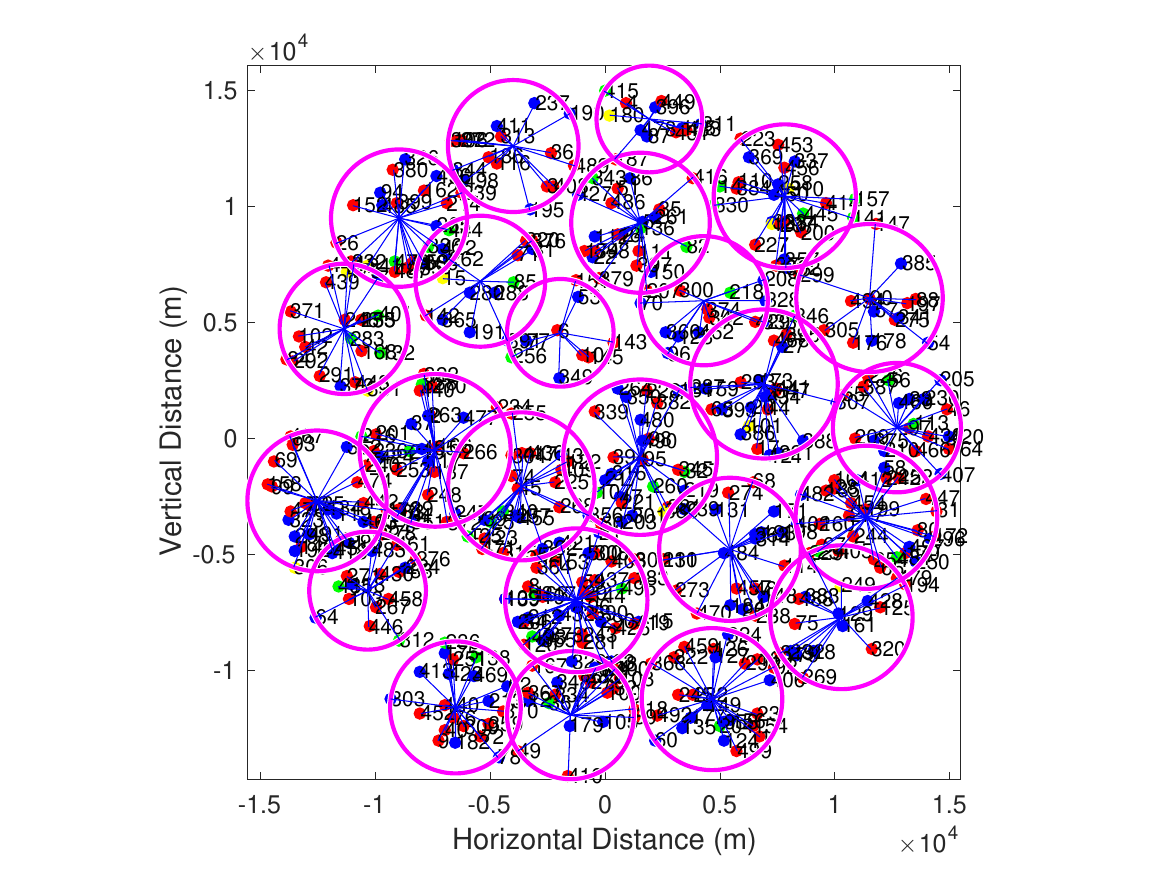}} \hspace{-1cm}
\subfigure[UAVB-NIBs: Optimal Altitudes]{\label{fig1:OA}\includegraphics[width = 0.35\linewidth]{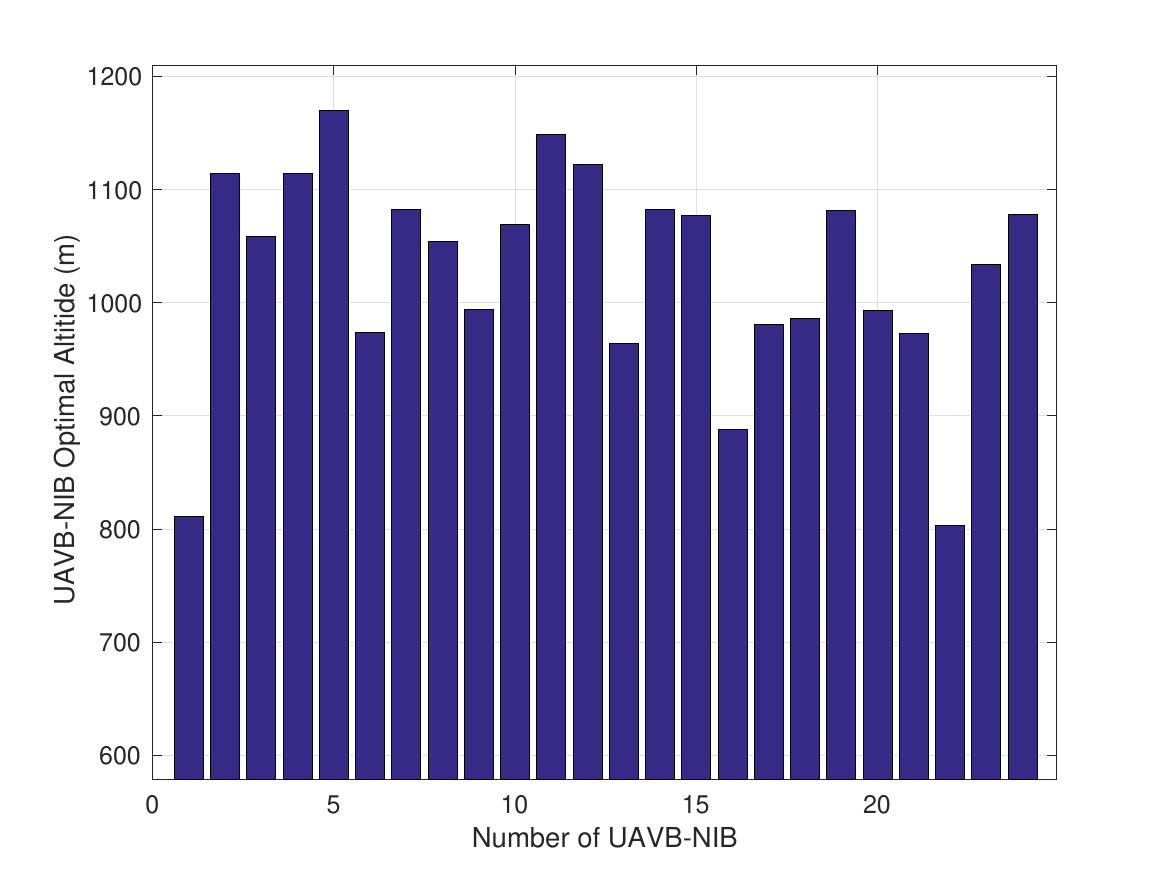}} \hspace{-0.84cm}
\subfigure[UAVB-NIBs: Backhaul Links ]{\label{fig:BL}\includegraphics[width=0.36\linewidth]{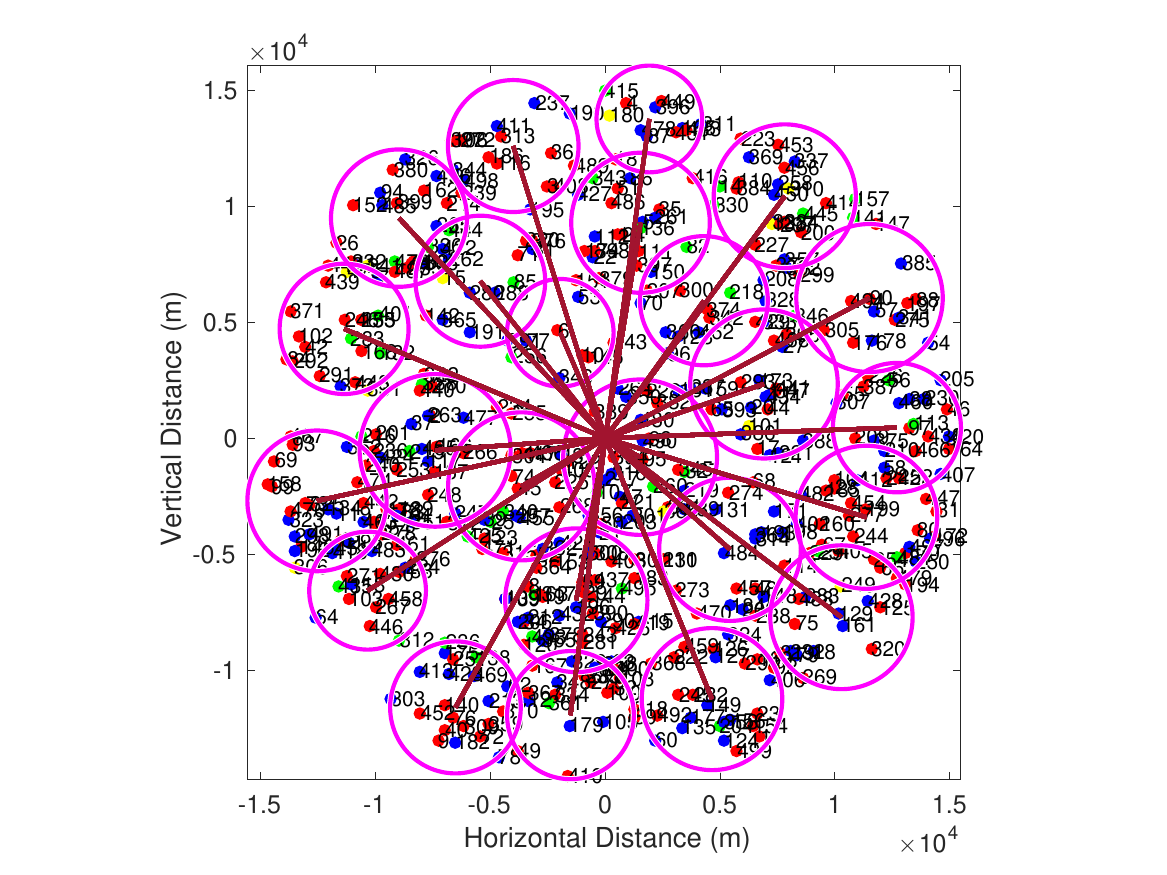}} 
\caption{UAVB-NIBs deployment, user association and parameters optimization}
\label{fig:UGA}
\end{figure*}
\section{Numerical Results and Discussion}
We adopt PHASA-$35$ HAPS aircraft model flying at an altitude of $20$km and evaluate the available transmit power for noon on the winter and summer solstice of 2025 using the solar algorithms \cite{javed2023interdisciplinary}. The analysis assumes different areas; sub urban, urban, dense urban, and high rise urban with different user densities, where users are distributed as Poisson Point Process. In addition, each UAVB-NIB is capable of offering $4$ RAT technologies to the GUs with different probabilities according to the user population, as illustrated in Fig. \ref{fig:RAT}. We assume distinct carrier frequencies and appropriate bandwidths for each RAT technology. We aim to carry out a thorough numerical analysis to quantify the gains achieved by each step in the proposed heterogeneous communication system. 

\subsection{Illustration of the Proposed Sequential Algorithm}
The sequential optimization of the proposed UAVB-NIB deployment, \ac{UA}, and location optimization is illustrated in Fig. \ref{fig:UGA}. Color-coded users operating at different RATs and distributed as Poisson point process in the circular HAPS coverage area of $150$km radius are demonstrated in Fig. \ref{fig1:Users}.RAT-1(yellow) is the least probable whereas RAT-4(blue) is the preferred access technology for the GUs in this configuration area using the probabilities in Fig. \ref{fig:RAT}. Next, we determine the initial coverage zones for UAVB-NIB deployment using the GDC algorithm, which render $24$ required UAVB-NIBs and their central locations for a given beam radius as depicted in Fig. \ref{fig1:GDC}. It is followed by the \ac{UA} using greedy algorithm where the users associate themselves with the UAVB-NIB based on the maximum received SINR.  Fig. \ref{fig1:UA} presents the \ac{UA} using green lines merging towards the beam center marked by the UAVB-NIB's projection on ground. 
This \ac{UA} allows us to concentrate the antenna beams towards the active users while optimizing the beams parameters and UAVB-NIBs locations. Fig. \ref{fig:BO} exhibits the optimal UAVB-NIB locations and reduced coverage/beam radii to serve the predefined associated users. Clearly, this minimizes the overlapping regions and maximizes the power density in a given beam which increases the SINR and consequently the sum-rate of the users. On the other hand, optimal flying altitudes of the $24$ participating UAVB-NIBs are highlighted in Fig. \ref{fig1:OA}. Eventually, Fig. \ref{fig:BL} displays the backhaul connection of each UAVB-NIB with the HAPS positioned at the origin without loss of generality. 
Fig. \ref{fig:UGA} displays the step-by-step approach to tackle the 
UAVB-NIB deployment and \ac{UA} problems. These optimized parameters can then be utilized for optimal resource allocation by UAVB-NIBs and HAPS in the access and backhaul downlink, respectively. 
 \begin{figure*}[htp]
  \centering
\subfigure[Proposed UAVB-NIB Deployment]{\label{fig:J_CR}\includegraphics[width=0.35\linewidth]{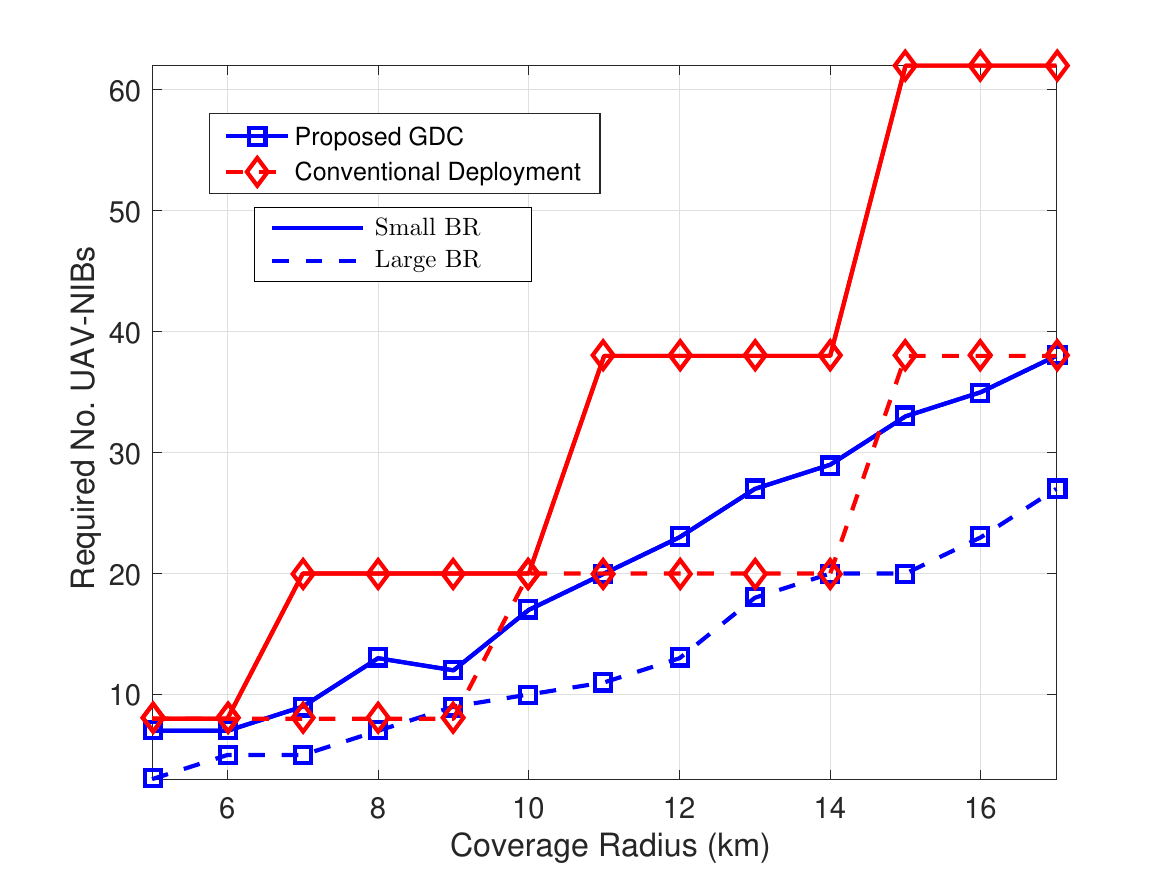}} \hspace{-0.83cm}
\subfigure[Different user association schemes]{\label{fig:SINR_Pt}\includegraphics[width = 0.35\linewidth]{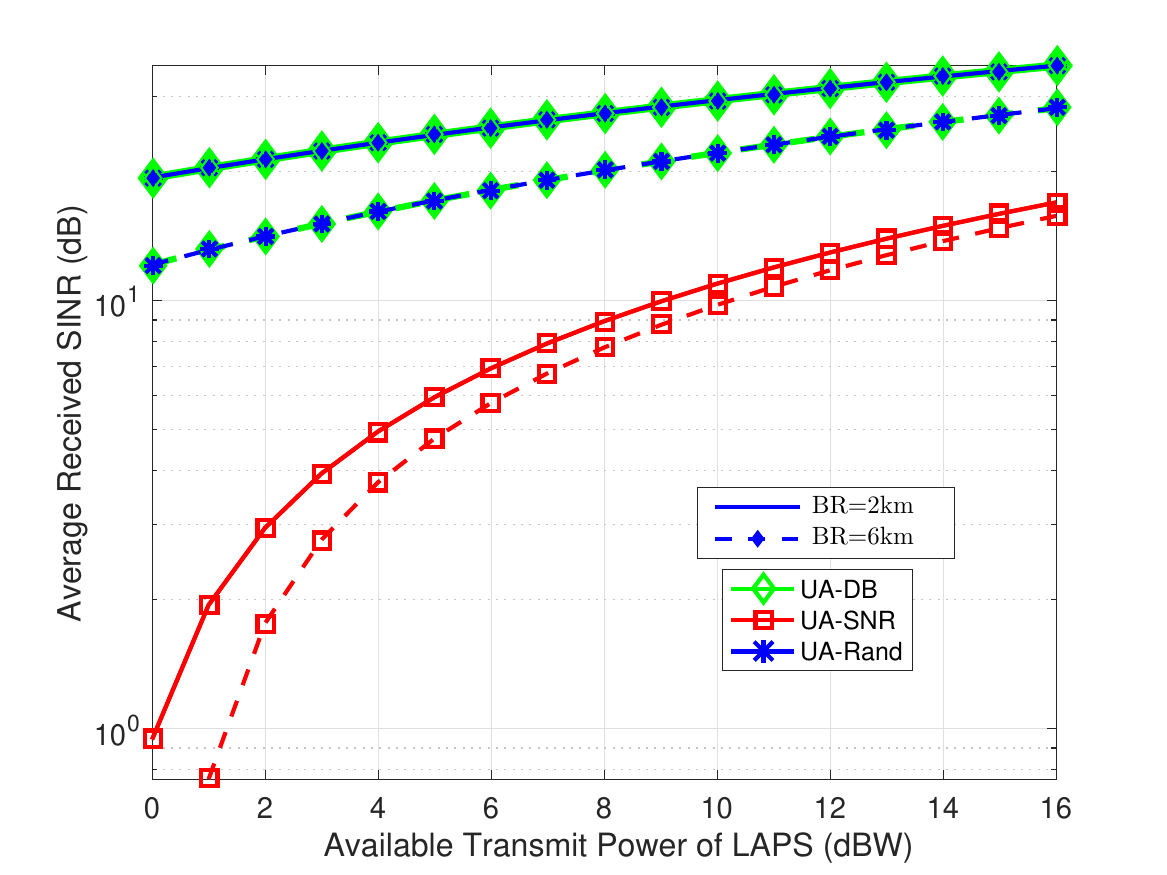}} 
\hspace{-0.84cm}
\subfigure[AEE and ASE of Backhaul]{\label{fig:AEE_ASE_HAPS}\includegraphics[width=0.35\linewidth]{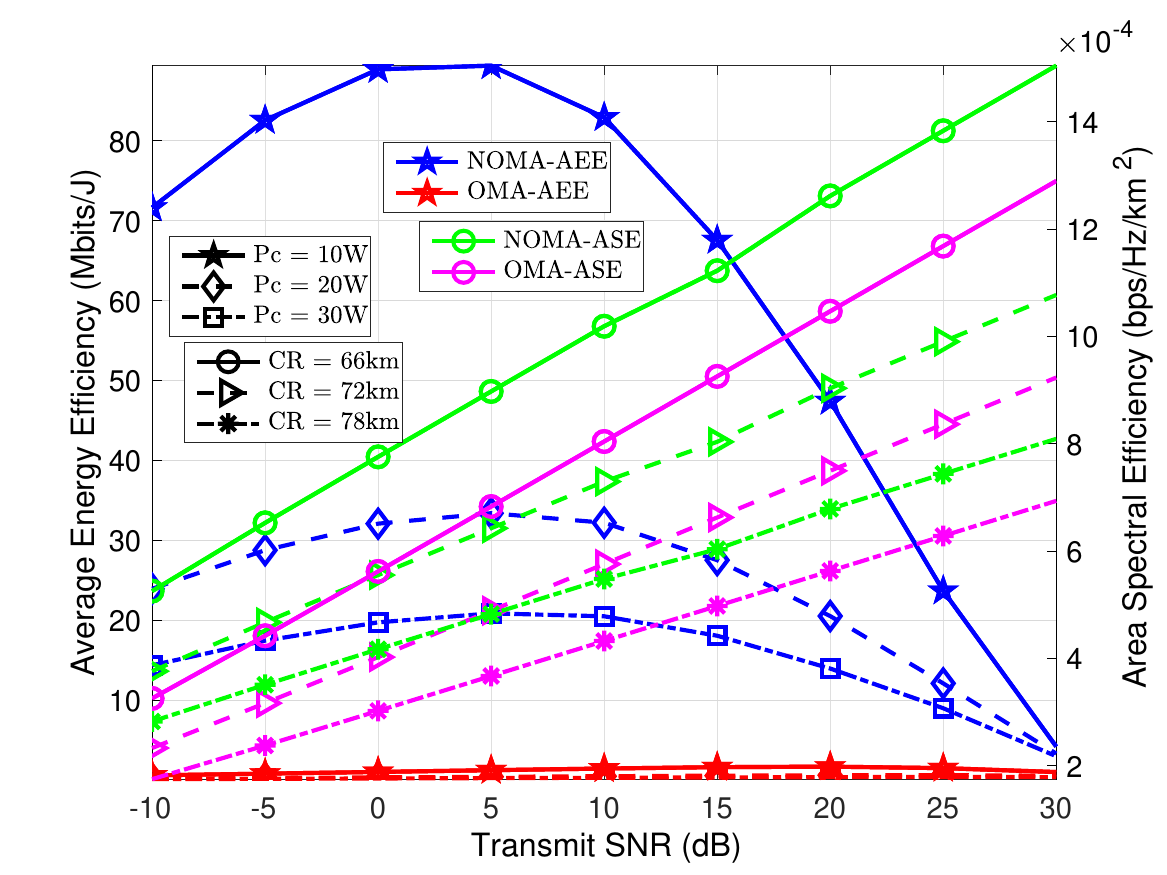}} 
\caption{Parameters optimization and resource allocation}
\label{fig:PO}
\end{figure*}
\subsection{UAVB-NIB Deployment Schemes}
In order to validate the effectiveness of UAVB-NIB deployment algorithm, we study the minimum number of required UAVB-NIBs in a sub-urban area with user density $1000$ users/$\rm{km}^2$ and radius ranging from $5$km to $17$km. We assume two different antenna beam sizes for the participating UAVB-NIBs termed as small and large beam-radii with values $2.5$km and $3.5$km, respectively. Evidently, the number of required UAVB-NIBs increases with the increasing coverage radius and consequently increasing number of GUs in that area as detailed in Fig. \ref{fig:J_CR}. Intuitively, a large number of narrow-beamed UAVB-NIBs are required to serve a certain coverage area as compared to the wide-beamed UAVB-NIBs. Interestingly, the proposed UAVB-NIB deployment GDC scheme renders significantly less number of required UAVB-NIBs as opposed to the conventional deployment. The step-wise increase in the conventional cellular deployment is owing to the increase in tier of UAVs for a certain change in coverage radius which then stays the same for a range of coverage radii before adding another tier/layer of UAVs. At around $16{\rm km}$ coverage radius the difference between proposed GDC and conventional approach is around $35$ narrow-beamed UAVB-NIBS as opposed to $62$ narrow-beamed UAVB-NIBS and $23$ wide-beamed UAVB-NIBS relative to the $38$ wide-beamed UAVB-NIBS, respectively. We can conclude that the proposed users-aware deployment scheme can serve a given coverage area with up to $50\%$ less UAVB-NIBs than the conventional deployment. 
\subsection{Performance of Different UA Strategies}
We evaluate the performance of the proposed SNR-based \ac{UA} (UA-SNR) algorithm with distance-based \ac{UA} algorithm (UA-DB) and random \ac{UA} (UA-Rand) in Fig. \ref{fig:SINR_Pt}. We investigate the average received SINR at the GUs in the same sub urban setting with narrow-beamed and wide-beamed UAVB-NIBs. For a given user distribution, we evaluate the optimal number and locations of the UAVB-NIBs. Then, we perform user-association with pre-defined UAVB-NIBs locations using the three aforementioned UA algorithms to study the average received SINR of the GUs. Expectedly, the average received SINR increases with the increased transmission power and especially for the narrow-beamed UAVB-NIBs. Interestingly, the UA-DB performs equally well as the proposed UA-SNR whereas UA-Rand demonstrates degraded performance for the entire range of transmission power. Using $10$dBW transmission power of each particpating UAVb-NIB, we observe almost $20$dB difference in the average received SINR of the GUs signifying the importance of the appropriate UA instead of random user association. 
\subsection{Energy and Spectral Efficiency of the Backhaul}
Aerial communications are predominantly restricted by the available power budget \cite{javed2023interdisciplinary}. This manifests energy  efficiency\footnote{EE is measured in bits/Joules i.e., a higher value of EE indicates the higher amount of data in bits that can be sent with minimal energy consumption.} as a critical performance metric for the solar-powered HAPS. Energy efficient communication with cognizant power control can significantly impact and prolong the flight operation times. 
We can compute the average energy efficiency (AEE) of the backhaul link 
${\rm AEE}_{\rm b}$ using
\begin{equation}
{\rm AEE}_{b} =   \frac{1}{J} \sum_{j=1}^{J}   \frac{R_{j}}{f_{j} P_{H} + P_{c_2}  }, 
\end{equation}
where $R_{j}$ is the achievable rate of $j^{\rm th}$ UAVB-NIB with power allocation $f_{j} P_{H}$ and circuit power consumption of $P_{c_2}$ in the backhaul.
Similarly, the spectrum efficiency\footnote{It is a measure of how efficiently a limited frequency spectrum is utilized to transmit the data by the proposed communication system. It is typically measured in bps/Hz.} describes the amount of data transmitted over a given spectrum with minimum transmission errors. The average SE of the backhaul link with NOMA can be viewed as ${\rm SE}_{\rm avg}^{b} =  \frac{1}{J} \sum_j {R_{j}/B_{H}}$, while the area SE (ASE) of the backhaul link can be presented~as:
\begin{equation}
{\rm ASE}_b =  \frac{{\rm SE}_{\rm avg}^{b} }{\pi R^2},
 \end{equation}
The described AEE and ASE of the backhaul link between HAPS and UAVB-NIBs are analyzed for optimal versus sub-optimal resource allocation strategies in Fig. \ref{fig:AEE_ASE_HAPS}. We assume an urban area of $60$km radius and $3000$users/$\rm{km}^2$ user density served by the wide-beamed UAVB-NIBs with beam radii $5$km. We further assume $6.4$GHz carrier frequency with $100$MHz channel bandwidth for the backhaul connection. The AEE is observed for the range of transmit SNR for three different circuit power consumption scenarios i.e., Pc$=10$W, Pc$=20$W, and Pc$=30$W.
Evidently, the AEE of NOMA based power allocation outperforms the OFDMA counterpart for the entire range of transmit SNR. Moreover, the NOMA-AEE decreases with increasing circuit power consumption whereas the NOMA-AEE initially increases with increasing transmit SNR and then decreases with further increase in transmit SNR. This renders the maximum NOMA-AEE of $90$Mbps/J around $5$dB transmit SNR for all circuit power consumptions.  In addition, the Fig. \ref{fig:AEE_ASE_HAPS} reveals the ASE of the same system for three different coverage radii. Clearly, the ASE decreases with increasing coverage area as the same HAPS resources are now utilized to serve increasing number of users requiring more UAVB-NIBs with fixed beam coverage areas. Moreover, the SE increases with the increasing transmit SNR owing to the increase in the achievable rate. Lastly, the NOMA-ASE outperforms OMA-ASE for all transmit SNR and all coverage radii. For instance, NOMA-ASE renders around $20\%$ percentage increase over conventional OMA-ASE at $20$dB transmit SNR. 
 \begin{figure*}[htp]
  \centering
\subfigure[Average sum rate of AL]{\label{fig:SumR_Pt}\includegraphics[width=0.33\linewidth]{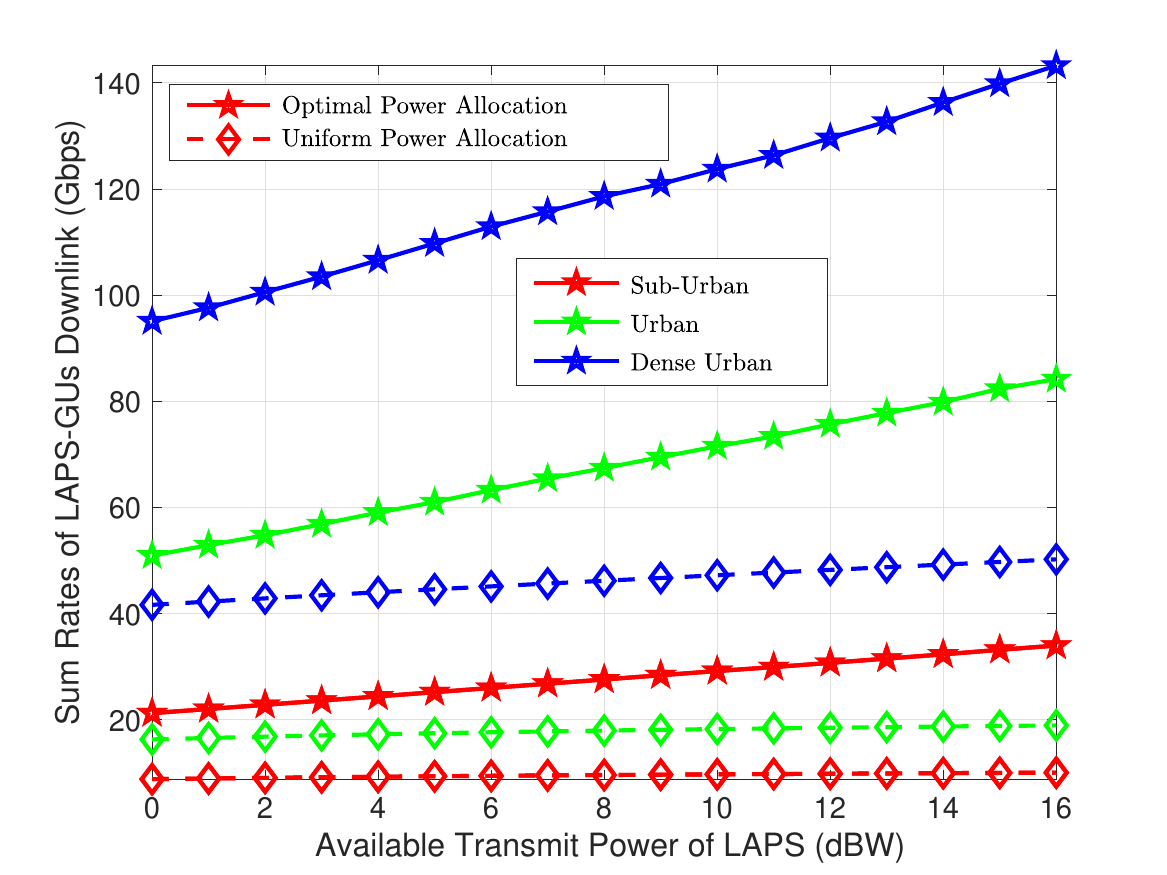}} \hspace{-0.78cm}
\subfigure[Average fairness index]{\label{fig:AFI_SNR}\includegraphics[width = 0.34\linewidth]{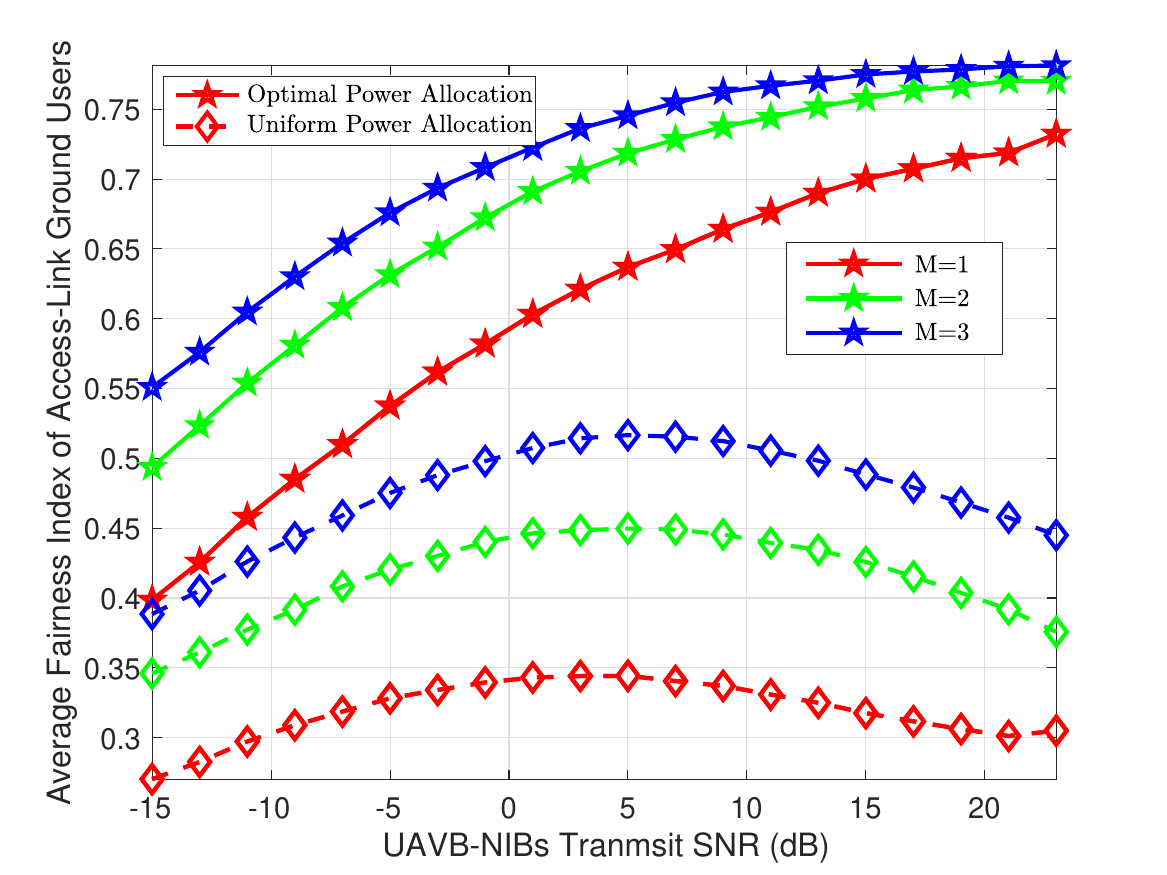}} 
\hspace{-0.82cm}
\subfigure[AEE and ${\rm SE}_{\rm avg}$ of AL]{\label{fig:AEE_ASE_LAPS}\includegraphics[width=0.34\linewidth]{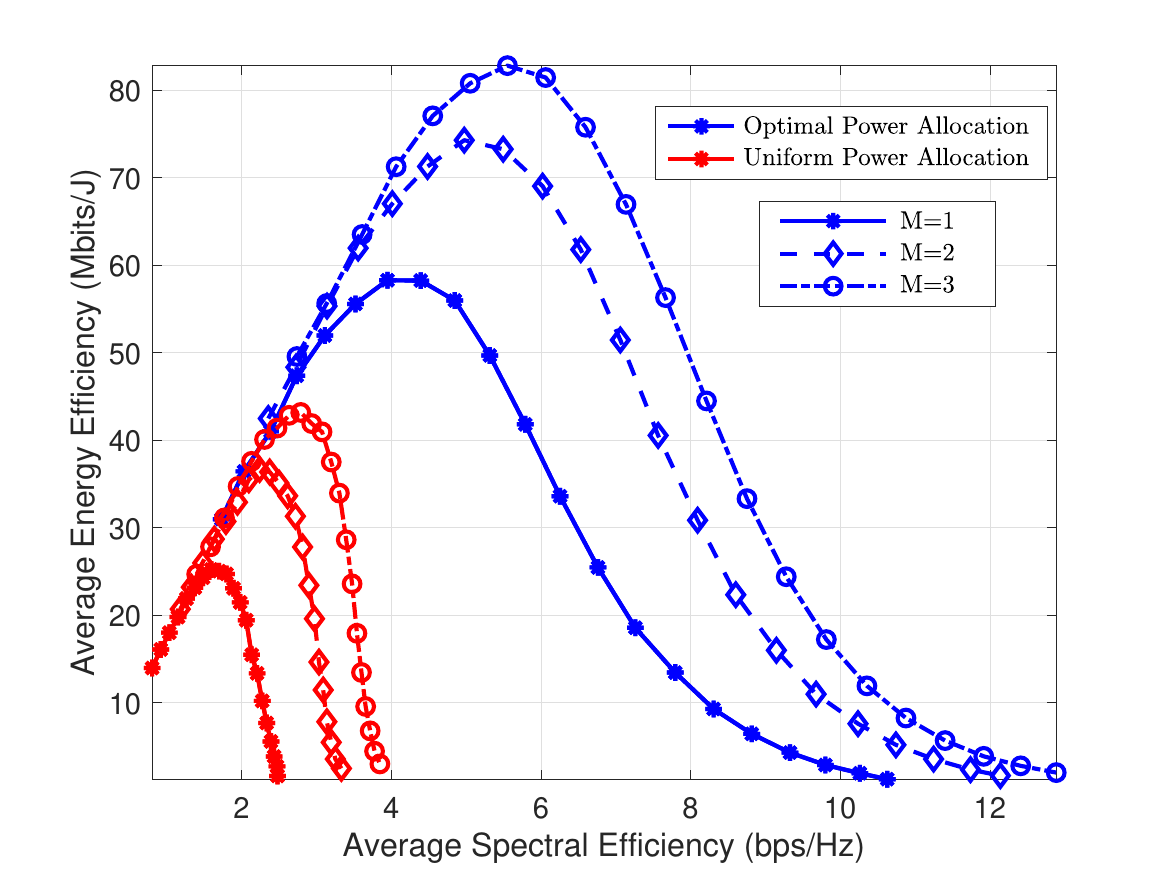}} 
\caption{Effect of optimal resource allocation on sum-rate, fairnes index, EE and SE in the AL}
\label{fig:AL}
\end{figure*}
\subsection{Average Sum Rate Performance}
The impact of the proposed non-uniform power allocation (NUPA) strategy is compared with the conventional uniform power allocation (UPA) on the average sum rate of the AL.  Fig. \ref{fig:SumR_Pt} presents the downlink sum rate with the UAVB-NIB transmission power budget ranging from $0$dBW to $16$dBW for three different area configurations. Sub urban, urban, and dense urban areas are assumed to have distinct user densities i.e.,$1000$users/$\rm{km}^2$, $3000$users/$\rm{km}^2$, and $5000$users/$\rm{km}^2$, respectively. In addition, they have unique environmental parameter pairs ($\eta_{\rm LOS},\eta_{\rm NLOS}$);  $(0.1, 21), (1.0,20)$, and $(1.6, 23)$ in respective order \cite{al2014modeling}. The values of the constants are $a = 11.95$, and $b =0.136$ \cite{sun2019user} whereas the RAT operational demands according to the area configurations are shown in Fig. \ref{fig:RAT}.
We suppose maximum antenna gain of UAVB-NIBs $G_{\rm max} = 23$dBi, HPBW $\theta_J^{\rm 3dB} = 12^o$ and transmit diversity $M=2$, unless specified otherwise. The sum rates of LAPS-GUs downlink are averaged over numerous channel instances.  Evidently, the sum rate increases with the increase in transmission power, however, the gain is particularly significant in optimal power allocation strategy. We further observe that dense urban area exhibits highest sum rates owing to the high user-density requiring concentrated beams from a large number of UAVB-NIBs to serve the given coverage area. We can observe the percentage increase in average sum rate upto $185.24\%$ with the proposed optimal power allocation over the uniform power assignment. 
\subsection{Jain's fairness index}
The fairness of a communication system is analyzed to determine whether all participating nodes are receiving a fair share of the system resources.  The UF of the GUs in the AL can be quantified using the Jain's fairness index as:
 \begin{equation}
 {\mathcal {J}}_a={\frac {(\sum _{k=1}^{K}R_{kj}^{\Omega})^{2}}{K \cdot \sum _{k=1}^{K}{R_{kj}^{\Omega}}^{2}}}.
  \end{equation}
The UF is evaluated for a dense urban area versus transmit SNR for varying number of transmit antennas in Fig. \ref{fig:AFI_SNR}. The average fairness index (AFI) is the fairness index of all users in the given coverage area averaged over numerous channel instances. Intriguingly, the AFI increases with increasing transmit diversity and optimal power allocation clearly renders higher AFI than the uniform counterpart. Interestingly, we observe different trends of the two power allocation schemes with respect to the transmit SNR. Increasing transmit 
SNR results in increasing fairness amongst users for the NUPA. However, it initially increases and then decreases with increasing transmit SNR for the UPA. The UPA yields maximum AFI at $5$dB SNR but this is still $30\%$ less than the AFI of optimal power allocation at same SNR. We observe $85\%$, $59.75\%$ , and $44.28\%$ percent improvement in AFI using the proposed power allocation over unifrom power allocation with $M=1,2$, and $3$ transmit antennas, respectively. 
\subsection{System Efficiency of the AL}
The system efficiency of the AL can also be evaluated in terms of AEE and ASE. 
The EE metric is specifically beneficial for UAVB-NIBs operating in remote or hard-to-access areas where battery replacement or recharging may be challenging. In the AL, the AEE of the AL can be seen as the averaged EE of all the users under UAVB-NIBs coverage i.e., 
\begin{equation}
{\rm AEE}_{\rm a}= \frac{1}{K} \sum_{j=1}^{J} \sum_{k} \sum_{\Omega} \frac{R_{kj}^{\Omega}}{p_{kj}^{\Omega} P_{j}^{\Omega} + P_{c_1}  } \qquad \forall k,
\end{equation}
where $R_{kj}^{\Omega}$ and $p_{kj}^{\Omega}$ are the achievable rate and power allocation factors of the concerned user in the respective order, $P_{j}$ and $ P_{c_1}$ are the power budget and circuit power expenditures, respectively. Moreover, assuming the perfect CSI and RZF, we can write the average SE of the AL as: 
\begin{equation}
{\rm SE}_{\rm avg}^{a} =  \frac{1}{K} \sum_j \sum_{k} \sum_{\Omega} {R_{kj}^{\Omega}}/{B_{j}^{\Omega}} 
\end{equation}
The overall SE is averaged for all RATs as each one offers its own bandwidth to the operational set of users. The AEE versus ${\rm SE}_{\rm avg}^{a}$ of the AL is evaluated for the same system parameters. The AEE appears to be a concave function of ${\rm SE}_{\rm avg}^{a}$ rendering maximum value of $82.84$Mbits/J at $5$bps/Hz for optimal power allocation as opposed to the maximum value of $43.2$Mbits/J at $3$bps/Hz for UPA. Fig. \ref{fig:AL} demonstrates the concentrated values on the lower-left bottom of the graph for UPA as compared to the right-top values for optimal power allocation. This signifies the lower AEE and ${\rm SE}_{\rm avg}^{a}$ values of UPA versus higher AEE and ${\rm SE}_{\rm avg}^{a}$ values of optimal power allocation for any number of transmit antennas. We can quantify the percentage improvement in the peak AEE values with optimal power allocation over the UPA as $132.52\%$, $104.2\%$, and $91.77\%$ for $M=1,2,$ and $3$ transmitting UAVB-NIB antennas, respectively.  
\section{Conclusions}
The compact, portable, and versatile NIB solution along with heterogeneous aerial communication platforms i.e., LAPS and HAPS can formulate an intelligent aerial wireless network for remote coverage. Such combination for the access and backhaul link offers flexible, scalable, and resilient coverage solution. We have proposed NOMA for the backhaul connection between HAPS and UAVB-NIBS whereas zero-forcing scheme for the access MISO downlink channel between UAVB-NIBs and GUs. In addition, we have presented geometric disk cover for optimal UAVB-NIB deployment, greedy algorithm for user association, Lagrange optimization for the location optimization and successive convex approximation for the resource allocation problem in order to improve the system performance. The proposed algorithms and design guidelines enable this intelligent network to configure, deploy, and serve  with the  effective resource allocation, minimal power consumption, and enhanced system performance at the desired coverage area as per the users demand and QoS thresholds. Numerical results have revealed up to $50\%$ less number of required UAVB-NIBs, $20\%$ improvement in backhaul ASE, $185.24\%$ increase in the average sum rate of the AL, and $85\%$ improvement in average UF with the proposed strategies. 
\section{Acknowledgment}
This work is supported in part by the King Abdullah University of Science and Technology Research Funding (KRF) under Award ORA-2021-CRG10-4696. The work of Yunfei Chen is also supported by EPSRC TITAN (EP/Y037243/1, EP/X04047X/1).  {The work of Cheng-Xiang Wang is also supported by the National Natural Science Foundation of China (NSFC) under Grant 61960206006 and the EU H2020 RISE TESTBED2 project under Grant 872172.}
\appendices
\section{Optimal Elevation Angle}  \label{AppendA}
We can write the path loss of the AL between $k^{\rm th}$ user and $j^{\rm th}$ UAVB-NIB operationg at $\Omega$ RAT as a function of the elevation angle $\phi_{jk}^{\Omega}$ as:
\begin{equation}\label{eq.PL2}
L_{jk}^{\Omega}[{\rm dB}] = \frac{A}{1\!\! +\! \!{\bar a} e^{ {-b \phi_{jk}^{\Omega}}}} + 20 \log_{10} (r_j \sec \phi_{jk}^{\Omega} )+  \bar{B}^{\Omega}
\end{equation}
where ${\bar a} = a e^{ab}$ and
\begin{equation}
 \bar{B}^{\Omega} =  20 \log_{10} \left( \frac{4 \pi f_c^{\Omega}}{c}  \right) + \eta_{\rm NLOS}  
\end{equation}
Interestingly, for the worst case scenario $\phi_{jk}^{\Omega} = \phi_{j}$ i.e., path loss is maximum at the cell edge. Thus, solving \textbf{P1(c)} is equivalent to minimizing $L_{jk}^{\Omega}[{\rm dB}]$ with respect to $ \phi_{j}$ for the given $r_j^*$ and ${\bf w}_j^*$. 
It is straight forward to prove that the path loss in \eqref{eq.PL2} is convex in $\phi_{j}$ by showing that $\frac{\partial^2 L_{jk}^{\Omega}}{\partial \phi_j^2} \geq 0$. Hence, the first-order stationary point is the optimal  $\phi_{j}^*$, which can be attained by solving $\frac{\partial L_{jk}^{\Omega}}{\partial \phi_j} = 0$. 
 \bibliographystyle{IEEEtran}
\bibliography{IEEEabrv,refsNIB}

\end{document}